\documentclass[12pt]{article}
\usepackage{amsmath}
\usepackage{amsfonts}
\usepackage{amssymb}
\usepackage{amsthm}
\usepackage{cite}
\usepackage{hyperref}
\usepackage{mathrsfs}
\usepackage{subcaption}
\usepackage{cancel}
\usepackage{color}  
\usepackage{tikz}
\usepackage{graphicx}
\usepackage{soul}
\usetikzlibrary{decorations.pathreplacing}

\newcommand{\crr}{\color{black}}  

\newcommand{\rtable}{\color{black}} 
\newcommand{\rrtable}{\color{red}}
\newcommand{\gtable}{\color{green}}
\newcommand{\red}{\color{black}}

\usepackage{bbding}
\usepackage{comment}
\includecomment{comment}

\usepackage{mathrsfs}

\newlength{\dinwidth}
\newlength{\dinmargin}

\usepackage{tikz}
\usetikzlibrary{calc}
\usetikzlibrary{trees}
\usetikzlibrary{decorations.pathmorphing}
\usetikzlibrary{decorations.markings}
\usepackage{wrapfig}

\newcommand{\vv}{g}

\newcommand{\mrm}{\mathrm}

\renewcommand{\mathbf}{\boldsymbol}

\newcommand{\mcU}{\mathcal U}

\newcommand{\mcR}{\mathcal R}
\newcommand{\mcF}{\mathcal F}

\newcommand{\mcL}{\mathcal L}

\renewcommand{\i}{\mathrm i}

\newcommand{\ti}{\tilde}

\newcommand{\Om}{\Omega}

\newcommand{\La}{\Lambda}

\newcommand{\hh}{\hat h }

\newcommand{\vk}{\boldsymbol{k}}

\newcommand{\be}{\gamma}

\newcommand{\eps}{\varepsilon}
\newcommand{\de}{\delta}

\newcommand{\De}{\Delta}
\newcommand{\e}{\mathrm{e}}

\newcommand{\nin}{\noindent}
\newcommand{\si}{\sigma}

\newcommand{\h}{\fr{1}{2}}
\newcommand{\nat}{\mathbb{N}}

\newcommand{\om}{\omega}
\newcommand{\mfa}{\mathfrak{A}}
\newcommand{\mco}{\mathcal{O}}

\newcommand{\supp}{\mathrm{supp}}
\newcommand{\fr}[2]{\frac{#1}{#2}}
\newcommand{\al}{\beta}
\newcommand{\real}{\mathbb{R}}
\newcommand{\complex}{\mathbb{C}}

\newcommand{\non}{\nonumber}

\newcommand{\lan}{\langle}
\newcommand{\ran}{\rangle}

\def\proof{\noindent{\bf Proof. }}
\def\qed{$\Box$\medskip}

\newtheorem{theoreme}{Theorem } [section]
\newtheorem{proposition}[theoreme]{Proposition}
\newtheorem{lemma}[theoreme]{Lemma}
\newtheorem{definition}[theoreme]{Definition}
\newtheorem{corollary}[theoreme]{Corollary}
\newtheorem{remark}[theoreme]{Remark}
\newtheorem{example}[theoreme]{Example}
\newtheorem{criterion}[theoreme]{Criterion}
\newtheorem{conjecture}{Conjecture}
\newtheorem{assumption}{Assumption}

\newcommand{\bea}{\begin{assumption}}
	\newcommand{\eea}{\end{assumption}}
\newcommand{\beco}{\begin{conjecture} }
	\newcommand{\eeco}{\end{conjecture} }
\newcommand{\beq}{\begin{equation}}
	\newcommand{\eeq}{\end{equation}}
\newcommand{\beqa}{\begin{eqnarray}}
	\newcommand{\eeqa}{\end{eqnarray}}
\newcommand{\ben}{\begin{arabicenumerate}}
	\newcommand{\een}{\end{arabicenumerate}}
\newcommand{\bex}{\begin{example}}
	\newcommand{\eex}{\end{example}}
\newcommand{\ber}{\begin{remark}}
	\newcommand{\eer}{\end{remark}}
\newcommand{\bec}{\begin{corollary}}
	\newcommand{\eec}{\end{corollary}}
\newcommand{\bep}{\begin{proposition}}
	\newcommand{\eep}{\end{proposition}}
\newcommand{\becr}{\begin{criterion}}
	\newcommand{\eecr}{\end{criterion}}

\def\bel{\begin{lemma}}
	\def\eel{\end{lemma}}
\def\bet{\begin{theoreme}}
	\def\eet{\end{theoreme}}
\def\bed{\begin{definition}}
	\def\eed{\end{definition}}

\newcommand{\quasi}{\approx}

\newcommand{\T}{T}

\newcommand{\thet}{\r}

\newcommand{\hzeta}{\zeta^{(\r)}}
\newcommand{\stw}{}

\newcommand{\tfmal}{\ti f_{\al, \mu}}
\newcommand{\tfmbe}{\ti f_{\be,\mu}}

\newcommand{\tbfmal}{\ti{\bar{f}}_{\al, \mu}}
\newcommand{\tbfmbe}{\ti{\bar{f}}_{\be, \mu}}

\newcommand{\nI}{\#I}
\newcommand{\vertiii}[1]{{\left\vert\kern-0.25ex\left\vert\kern-0.25ex\left\vert #1 
    \right\vert\kern-0.25ex\right\vert\kern-0.25ex\right\vert}}

\newcommand{\taur}{\tau^{(\thet)}}

\newcommand{\Tr}{\T_{\thet}{}}

\newcommand{\2}{&\!\!\!}

\newcommand{\vac}{\mrm{vac} }

\newcommand{\mcW}{W}

\newcommand{\ww}{\hat{b}}
\renewcommand{\hh}{{ \fr{1}{4} }}

\renewcommand{\k}{\boldsymbol{k}}
\newcommand{\x}{\boldsymbol{x}}
\renewcommand{\v}{\boldsymbol{v}}
\renewcommand{\r}{r}
\newcommand{\Psir}{\Psi^{(r)}}
\newcommand{\mcUr}{\mcU_{\r}}
\newcommand{\pn}{(i_0-1)}
\newcommand{\xn}{a}
\newcommand{\xii}{\zeta^{\circ}}

\begin{document}
\title{Quantum phase transition of infrared radiation} 

\author{Bartosz Biadasiewicz and Wojciech Dybalski\\\\
Faculty of Mathematics and Computer Science \\  
Adam Mickiewicz University in Pozna\'n\\
ul. Uniwersytetu Pozna\'nskiego 4, 61--614 Pozna\'n, Poland.\\
\small{E-mails: {\tt bartosz.biadasiewicz@amu.edu.pl,  wojciech.dybalski@amu.edu.pl. }} \\\\
}
\date{}
\maketitle
\begin{abstract} 
We describe a phase transition of infrared radiation, driven by quantum fluctuations, which takes place at the  boundary of  (the conformal diagram of) Minkowski spacetime.  
Specifically, we consider a family of states interpolating between the vacuum 
and the Kraus-Polley-Reents infravacuum. A state from this family can be imagined as  a  static source emitting  flashes of infrared radiation in distant past. The flashes are in suitable squeezed states and the time intervals between them are controlled by a certain parameter $r$. For $r<0$ the states are lightcone normal, thus physically
indistinguishable from local excitations  of the  vacuum. They suffer from the
usual infrared problems such as  disintegration of the Bloch-Nordsieck $S$-matrix and
rotational symmetry breaking by  soft photon clouds. However, for $r>0$ lightcone normality
breaks down, the $S$-matrix is stabilized by the Kraus-Polley-Reents mechanism and
the rotational symmetry is restored. We interpret these two situations as  ordered ($r<0$) and disordered ($r>0$) phase of infrared radiation, and show that they can be distinguished by  asymptotic fluctuations of the fields. We also determine the singular behaviour of some $S$-matrix elements near the critical point $r=0$.

\end{abstract}

\newpage

\tableofcontents
\section{Introduction}
\setcounter{equation}{0}

Phase transitions are abundant in Nature and  belong to the most studied phenomena in  Physics \cite{So11, Zi}.
In spite of tremendous theoretical progress, especially pertaining to renormalization and
universality, they seem to escape any definite classification. Without any intention of reviewing
the subject, let us give some examples illustrating this diversity:   On the one hand, in  the textbook case 
of the ferromagnetic phase transition the relevant states are in thermal equilibrium and the transition is
driven by thermal fluctuations. {\rtable On the other hand,} quantum phase transitions typically occur at zero temperature, the relevant
states are ground states,  and the transitions  are caused by quantum fluctuations \cite{Vo03}. In the same time, 
the concept of phase transitions is not reserved for ground states and thermal states - there is a rich field of non-equilibrium
phase transitions with an important example of directed percolation \cite{Hi06}. To broaden the perspective,
it should also be noted that the bulk and the boundary of a physical system may have distinct phase diagrams \cite{Di97}. 
In particular the boundary may undergo a phase transition even if the bulk is far from  criticality \cite{KSMLSC22}. 

The boundary degrees of freedom of massless QFT enjoyed a lot of attention over the last decade.
In fact, the Strominger's `infrared triangle'  \cite{HMPS14, HPS16, St17, Pa17}  linking  the Weinberg's soft photon theorem, asymptotic 
charges and memory effects  is located at  the boundary of the (conformal diagram of)  Minkowski spacetime. All the vertices of this `infrared triangle' 
have a long history  \cite{We65, AS81, Br77, Sta81,He95, BG13} summarized in \cite{He16}, but the emergence of the triangle clearly intensified efforts to understand the asymptotic structure of electrodynamics and gravity, see e.g. \cite{CL15, GS16, CE17,HIW16, KPRS17, DH19, DW19, RS20, PSW22, MRS22}.
In spite  of such activity,  to our knowledge, the phase diagram of boundary degrees of freedom in massless QFT has not been studied to date. 
In the present paper we demonstrate that phase transitions actually occur for {\red these} degrees of freedom, even in massless scalar free field theory.

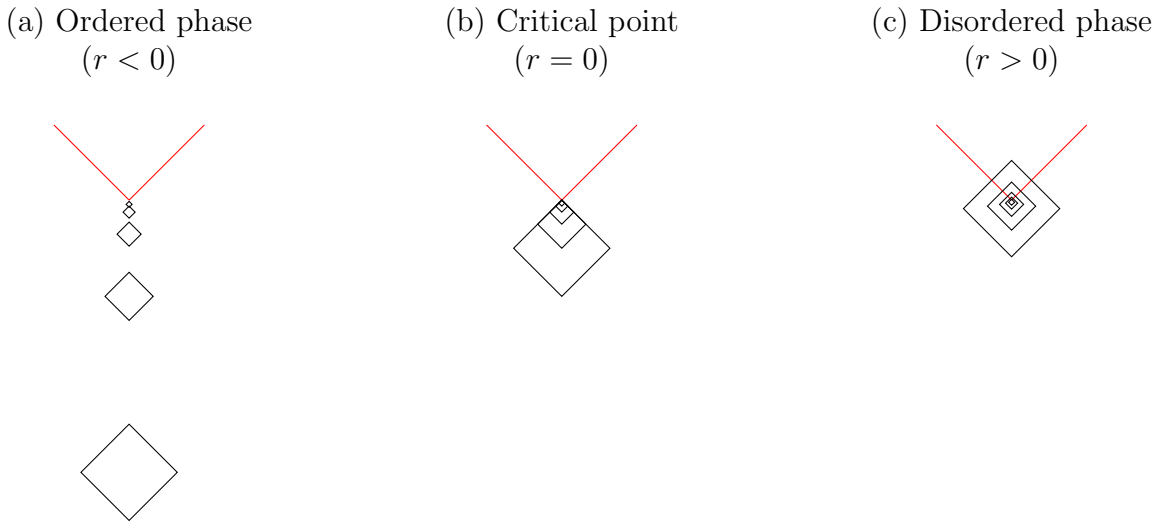
\begin{figure}[t]

\centering
\begin{tikzpicture}[scale=0.02][h]
	\draw (0,+100) node[anchor=south] {(a) Ordered phase};
	\draw (0,+75) node[anchor=south] {($r<0$)};
	\draw[red] (0,0) -- (50,50);
	\draw[red] (0,0) -- (-50,50);
	\pgfmathsetmacro{\r}{-1/2}
	
	\foreach \i in {2,3,4,5,6}{
		\pgfmathsetmacro{\epsilon}{pow(2, -\i + 1)}
		\pgfmathsetmacro{\translateY}{pow(\epsilon, -1 + \r)}
		
		\coordinate (A) at (0, 1/\epsilon);
		\coordinate (B) at (1/\epsilon, 0);
		\coordinate (C) at (0, -1/\epsilon);
		\coordinate (D) at (-1/\epsilon, 0);
		
		\coordinate (A') at ($(A) + (0,-\translateY)$);
		\coordinate (B') at ($(B) + (0,-\translateY)$);
		\coordinate (C') at ($(C) + (0,-\translateY)$);
		\coordinate (D') at ($(D) + (0,-\translateY)$);
		
		\draw[fill=none] (A') -- (B') -- (C') -- (D') -- cycle;
	}
	\draw[white] (0,-220) -- (0,-221);
\end{tikzpicture}
\hspace{2cm}
\begin{tikzpicture}[scale=0.02][h]
	\draw (0,+100) node[anchor=south] {(b) Critical point};
	\draw (0,+75) node[anchor=south] {($r=0$)};
	\draw[red] (0,0) -- (50,50);
	\draw[red] (0,0) -- (-50,50);
	\pgfmathsetmacro{\r}{0}
	
	\foreach \i in {2,3,4,5,6}{
		\pgfmathsetmacro{\epsilon}{pow(2, -\i + 1)}
		\pgfmathsetmacro{\translateY}{pow(\epsilon, -1 + \r)}
		
		\coordinate (A) at (0, 1/\epsilon);
		\coordinate (B) at (1/\epsilon, 0);
		\coordinate (C) at (0, -1/\epsilon);
		\coordinate (D) at (-1/\epsilon, 0);
		
		\coordinate (A') at ($(A) + (0,-\translateY)$);
		\coordinate (B') at ($(B) + (0,-\translateY)$);
		\coordinate (C') at ($(C) + (0,-\translateY)$);
		\coordinate (D') at ($(D) + (0,-\translateY)$);
		
		\draw[fill=none] (A') -- (B') -- (C') -- (D') -- cycle;
	}
	\draw[white] (0,-220) -- (0,-221);
\end{tikzpicture}
\hspace{2cm}
\begin{tikzpicture}[scale=0.02][h]
	\draw (0,+100) node[anchor=south] {(c) Disordered phase};
	\draw (0,+75) node[anchor=south] {($r>0$)};
	\draw[red] (0,0) -- (50,50);
	\draw[red] (0,0) -- (-50,50);
	\pgfmathsetmacro{\r}{+1/2}
	
	\foreach \i in {2,3,4,5,6}{
		\pgfmathsetmacro{\epsilon}{pow(2, -\i + 1)}
		\pgfmathsetmacro{\translateY}{pow(\epsilon, -1 + \r)}
		
		\coordinate (A) at (0, 1/\epsilon);
		\coordinate (B) at (1/\epsilon, 0);
		\coordinate (C) at (0, -1/\epsilon);
		\coordinate (D) at (-1/\epsilon, 0);
		
		\coordinate (A') at ($(A) + (0,-\translateY)$);
		\coordinate (B') at ($(B) + (0,-\translateY)$);
		\coordinate (C') at ($(C) + (0,-\translateY)$);
		\coordinate (D') at ($(D) + (0,-\translateY)$);
		
		\draw[fill=none] (A') -- (B') -- (C') -- (D') -- cycle;
	}
	\draw[white] (0,-220) -- (0,-221);
\end{tikzpicture}
\caption*{\footnotesize{Fig.~1. Schematic illustration of the phase transition in terms of the double cones from Fig.~2. The 
future lightcone with tip at the origin is indicated in red.}}
\label{transition-figure}
\end{figure}

\subsection{Phase transition on Minkowski spacetime} \label{description-transition}

{\red Let us explain in non-technical terms how the transition we {\red found} comes about, leaving the detailed construction of the states to Section~\ref{KPR-section}. Let us imagine a stationary source
 emitting flashes of infrared radiation in distant past. Specifically, suppose that the  corresponding wave packets have
 widths $\eps_i>0$ in momentum space, where  the sequence $\{\eps_i\}_{i\in\nat}$ converges to zero. One such emission event is depicted in Fig.~2. At the moment of emission the wave packet is essentially localized in space}
 \begin{wrapfigure}{l}{0.25\textwidth}
\begin{tikzpicture}[scale=0.6][h]
	\draw (0,0) -- (2,2);
	\draw (2,2) -- (0,4);
	\draw (0,0) -- (-2,2);
	\draw (-2,2) -- (0,4);
	\draw[dashed] (-2,2) -- (-4,4);
	\draw[dashed] (0,4) -- (-2,6);
	\draw[dashed] (2,2) -- (4,4);
	\draw[dashed] (0,4) -- (2,6);
	\draw [->, >=stealth, decorate, decoration={snake, amplitude=.4mm, segment length=2mm, post length=1mm}] (1,4) to (2.5,5.5);
	\draw [->, >=stealth, decorate, decoration={snake, amplitude=.4mm, segment length=2mm, post length=1mm}] (1.5,3.5) to (3,5);
	\draw [->, >=stealth, decorate, decoration={snake, amplitude=.4mm, segment length=2mm, post length=1mm}] (2,3) to (3.5,4.5);
	\draw [->, >=stealth, decorate, decoration={snake, amplitude=.4mm, segment length=2mm, post length=1mm}] (-1,4) to (-2.5,5.5);
	\draw [->, >=stealth, decorate, decoration={snake, amplitude=.4mm, segment length=2mm, post length=1mm}] (-1.5,3.5) to (-3,5);
	\draw [->, >=stealth, decorate, decoration={snake, amplitude=.4mm, segment length=2mm, post length=1mm}] (-2,3) to (-3.5,4.5);
	\draw[->] (0,2) -- (0,3.5);
	\draw[->] (0,2) -- (1.5,2);
	\draw (0,2) -- (0,1.5);
	\draw (0,2) -- (-0.5,2);
	\draw (-0.21,+2.75) node[anchor=south] {{\rtable t}};
	\draw (+1.25,+1.30) node[anchor=south] {{\rtable\textbf{x}}};
\end{tikzpicture}
\caption*{\footnotesize{Fig.~2. Radiation emitted from a base of a  double cone.}}
\label{double-cone-figure}
\end{wrapfigure}
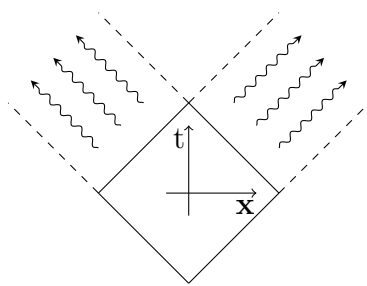
{\red  in a ball of radius $1/\eps_i$.   
This ball is a base of a double cone\footnote{A double cone of radius $a$ centered at zero is 
the open set $\mco_a:=\{\, (t,\x)\in \real^4\,|\, |t|+|\x|<a  \, \}$.}  in Fig. 2 which serves to determine
the region of spacetime to which the radiation will travel.
The $i$-th flash  occurs at time $-\taur_i$, where  }
\beqa
{\red \taur_i:= \left( \fr{1}{\eps_i}\right)  \eps_i^{\r}},     \label{taur-i}
\eeqa  
{\red and  $\r \in \real$ is our control parameter.  {\red Thus for $r=0$ this time-shift is equal to the radius of the double cone, as illustrated in  Fig. 1 (b).}
In  Fig. 1 (a) we presented a family of such double cones for $\taur_i$ given by
(\ref{taur-i}) and $\r$    negative, slightly smaller than zero. {\red Here the time-shifts are larger than the radii of the double cones. Consequently,} the double cones are located 
one below the other and  the infrared radiation evades large parts of spacetime, including the future lightcone.  
  On the other hand, in the case of   positive $\r$,  slightly larger than zero,  
we obtain a family of double cones inscribed within each other, as indicated in  Fig.~1~(c). Consulting again Fig.~2, we see that in this case the infrared radiation  fills the entire spacetime. The qualitative difference between these two situations  {\red (a) and (c)} suggests the presence of a phase transition at  $r=0$. We will show that this is, indeed, the case.}

To exhibit a phase transition it is necessary to specify order parameters. For this purpose, let us denote by $\phi$ the massless scalar
free field on Fock space $\mcF$ and by $\om_{\r}$ the physical state of radiation as discussed above. For a smooth, 
real valued function  $f$, we define, cf. \cite{Bu86},
\beqa
\phi_{R}(f):=\fr{1}{R^3} \int \phi(x) f(x/R)\, d^4x  \label{order-parameter}
\eeqa 
and choose  it as order parameter in the limit $R\to \infty$. We will be interested in two cases - $f$ supported in the spacelike complement
of zero  and in the future lightcone. We will refer to the respective quantities $\phi_R(f)$  as the \emph{spacelike} and \emph{timelike} order parameter.
Denoting by $\Om$ the Fock space vacuum, we obviously have 
$\lan \Om, \phi_{R}(f)\Om\ran=0$ and\footnote{We use the following  conventions for the Fourier transforms in space, time and spacetime:\\
$\tilde{f}(\k)=\fr{1}{(2\pi)^{3/2}} \int {\rtable d^3\x } \, \e^{-\i \k\cdot \x } f(\x)$, 
$\ti{f}(k^0)= \int dt\, \e^{\i k^0t} f(t)$, $\tilde{f}(k)
=\fr{1}{(2\pi)^{3/2}} \int  d^4x\, \e^{\i kx} f(x)$.} 
\beqa
\lan \Om,  \phi_{R}(f)^2 \Om\ran=\int \fr{{\rtable d^3 \k}}{2|\vk| }  |\ti{f}(|\vk|, \vk )|^2,
\eeqa
which is the rationale for the scaling chosen in (\ref{order-parameter}).  Since $\om_{\r}$ is a quasi-free state, we still have $\om_{\r}( \phi_{R}(f))=0$ for all $\r\in \real$
but the fluctuations  distinguish the two phases. Namely,
\begin{eqnarray}\label{distinction}
\lim_{R\to \infty}\om_{\r}(\phi_{R}(f)^2)= \left\{ \begin{array}{ll}
\int \fr{{\rtable d^3 \k}}{2|\vk| }  |\ti{f}(|\vk|, \vk )|^2 & \textrm{{\crr  for  $\r<0$,}}\\
\infty & \textrm{{\crr for $\r \geq 0$,}}
\end{array} \right.
\end{eqnarray}
where the divergence to infinity is possibly only along a subsequence and in both cases we will impose some restrictions on $f$.
In line with standard terminology we will call the region of parameters with bounded fluctuations the ordered phase
and the region with unbounded fluctuations the disordered phase. Since the fluctuations have quantum character and change in a discontinuous manner
with $\r$, we interpret this phenomenon as a quantum phase transition.   As the considered states are neither thermal nor
 vacua, the transition has a non-equilibrium character.  {\red It consists in a qualitative change of a certain dynamics, namely the Bloch-Nordsieck process, which we describe in the next subsection.}

\subsection{Spontaneous symmetry breaking} \label{spontaneous-symmetry-breaking}

The transition from the disordered to the ordered  phase is typically accompanied by spontaneous symmetry breaking. 
This phenomenon can be concisely described in the language of algebraic QFT. This  exact mathematical apparatus, proposed long time ago by Haag and Kastler \cite{HK64,Ha},  is  currently gaining popularity in theoretical high-energy physics \cite{Wi18, CLPW22, So23}. We find it 
very helpful for describing subtle properties of infrared radiation which are the topic of this paper. Denote by $\mfa$ the $C^*$-algebra generated by operators $\exp(\i \phi(f))$, where
$f\in D(\real^4;\real)$, i.e., a  smooth, compactly supported, real valued function on $\real^4$. Now $\om_{\r}$ is a state on $\mfa$, that is a positive, normalized, 
linear functional. We denote its  GNS representation by $\pi_{\r}$. Symmetries of our physical system are described by 
automorphisms $\alpha$ of $\mfa$. We say that a symmetry $\alpha$ is unitarily implemented in  representation $\pi_{\r}$ if there exists
a unitary $U$ such that $\pi_{\r}\circ \alpha(A)=U\pi_{\r}(A)U^*,  A\in \mfa$. In the usual shorthand notation
\beqa
\pi_{\r}\circ \alpha\simeq \pi_{\r}.  \label{implementation-of-unitaries}
\eeqa    
On the other hand, if such a unitary does not exist, we say that the symmetry $\alpha$ is broken in $\pi_{\r}$.

\newcommand{\phij}{\phi_{\mrm{int}}}

In {\rtable order} to find the relevant automorphisms $\alpha$, let us recapitulate the Bloch-Nordsieck discussion of infrared problems \cite{BN37}: Let  $\phij$ be the massless scalar quantum field interacting with an external source $j$
 of the form
\beqa\label{j-def}
j(t,\x)= \left\{ \begin{array}{ll}
j_0(\x-\v^{\mrm{out}}t)  & \textrm{for $t\geq 0$},\\
j_0(\x-\v^{\mrm{in}} t) & \textrm{for $t<0$},\\
\end{array} \right.
\eeqa
where $j_0\in D(\real^3;\real)$ is  spherically symmetric and satisfies $\int {\rtable d^3\x} \, j_0(\x)=:q\neq 0$. That is, the source corresponds to a
charged particle with  initial velocity $ \v^{\mrm{in}}$ and final velocity $\v^{\mrm{out}}$, both strictly smaller than the speed of light. 
The system is governed by  the field equation 
\beqa
 \Box \phij(x)=-j(x)
\eeqa  
with an initial condition $\phij(0,\x)=\phi_0(0,\x)$  which differs from $\phi$ at most by some real valued function. {\red (We cannot set $\phij(0,\x)=\phi(0,\x)$ here, since
we will identify $\phi$ with the incoming field below).}
Setting $V(t):=\int {\rtable d^3\x }\, \phi(t,\x)j(t,\x)$ we write  the textbook  solution for $t\geq 0$
\beqa
\phij(t,\x)=\mcW_t \phi_0(t,\x) \mcW_t^*, \quad \mcW_t:=\mrm{\bar{T}exp}\big(-\int_{0}^td\tau\, V(\tau)\big), \label{time-ordering}
\eeqa
where $\bar{T}$ orders the times $\tau$ ascendingly from left to right.
We thus identify the interacting dynamics as $U(t):=\mcW_t \e^{\i t H_0}$, where $H_0$ is the free Hamiltonian, and
compute the outgoing asymptotic field according to the LSZ prescription: 
\beqa
\phij^{\mrm{out}}(f)=\lim_{\tau\to \infty} U(\tau)\e^{-\i \tau H_0} \phi_0(f) \e^{\i \tau H_0} U(\tau)^*= \phi_0(f)-\i \mrm{Im}\lan \vv^{\mrm{out}},F\ran,  
\label{first-out}
\eeqa
where
\beqa
\vv^{\mrm{out}}(\k):= \i \ti j_0(\k) \,  \fr{1}{ |\vk|^{3/2}(1-\hat{\k} \cdot \v^{\mrm{out}}) },\quad\,  \hat{\k}:=\k/|\k|, \quad F(\k):=|\k|^{-1/2} \ti{f}(|\k|,\k).  \label{vv-def}
\eeqa
The latter definition is motivated by the relation $\phi(f)=\fr{1}{\sqrt{2}}(a^*(F)+a(F))$, where $a^{(*)}$ are the creation/annihilation operators\footnote{Our convention is
$[a(F_1), a^*(F_2)]=\lan F_1,F_2\ran$, where $\lan F_1,F_2\ran:=\int d^3{\rtable \k} \, \bar{F}_1({\rtable \k}) F_2({\rtable \k})$ is the scalar product {\red and 
$\|F\|_2=\lan F, F\ran^{1/2}$ the norm in} $L^2(\real^3)$.}
 on the Fock space $\mcF$. We repeat the computation for negative
times which amounts to changing the direction of time ordering in (\ref{time-ordering}) and replacing `out' with `in' in (\ref{first-out}) and (\ref{vv-def}).
Now we identify $\phi$ with the incoming field $\phij^{\mrm{in}}$ and define the automorphism $\alpha$ by the relation
\beqa
\e^{\i \phij^{\mrm{out}}(f)  }=\alpha(\e^{\i \phij^{\mrm{in}}(f)})= \e^{-\i \mrm{Im}\lan \vv^{\mrm{out}}-  \vv^{\mrm{in}}, F\ran }
 \e^{\i \phij^{\mrm{in}}(f)}. \label{alpha-automorphisms}
\eeqa
Comparing this relation with (\ref{implementation-of-unitaries}) it is clear that the implementing unitary $U$ (if it exists) is simply the
$S$-matrix for the  Bloch-Nordsieck process. {\red We will denote this unitary $S_r$.}

In fact, the $S$-matrix should intertwine the incoming and outgoing asymptotic fields. Its absence in the vacuum representation $\pi_{\vac}$ is the essence of the infrared
problem. Since $\lim_{\r\to{\crr -\infty}}\om_{\r}=\om_{\vac}$, i.e., our states tend weakly to the vacuum state for large {\crr negative $\r$}, it is 
not a surprise that the $S$-matrix  does not exist in the ordered phase of our system (see Section~\ref{desintegration-section}). As a matter of fact, it is also not a surprise
that  the Bloch-{\rtable Nordsieck} $S$-matrix  ${\red S_r}$ does exist in the disordered phase. {\red In fact}, we construct our states in such a way that    $\lim_{\r\to {\crr \infty}}\om_{\r}=\om_{\mrm{KPR}}$,
where $\om_{\mrm{KPR}}$ is the Kraus-Polley-Reents (KPR) infravacuum \cite{Re74,KPR77, Kr82, Ku98, CD18, BD22}. The latter state is known to have the implementation 
property (\ref{implementation-of-unitaries}) for the automorphisms in question as we recall in Section~\ref{Existence-S-matrix}.  It is thus clear from the fact that $\{\om_{\r}\}_{\r\in \real}$ interpolates between $\om_{\vac}$ and $\om_{\mrm{KPR}}$ that  the disintegration of {\red $S_{\r}$}  must occur for some intermediate value of $\r$.
We will show that it happens at $\r=0$ as a symmetry breaking phenomenon  accompanying the phase transition described in Subsection~\ref{description-transition}. 

As the symmetry transformation $\alpha$, discussed above, may appear somewhat abstract, let us have a brief look at the
familiar case of rotations. It turns out that our states $\om_{\r}$ are invariant under rotations for all $\r\in \real$, thus initially
there is no rotational symmetry breaking. The situation becomes more interesting if instead of $\om_{\r}$ we consider the 
family $\om_{\r}\circ \alpha$.  
In view of (\ref{alpha-automorphisms}), we simply transport our states from the algebra of incoming fields to the algebra of outgoing fields.
Thus we can interpret the resulting states as the background radiation from Fig.~1 and Fig.~2 accompanied by a soft photon cloud emitted by the external source.
In this case we have
\beqa
\2 \2\pi_{\r}\circ \alpha\circ \alpha_{\mcR}\not\simeq \pi_{\r}\circ \alpha\,\quad \textrm{for} \quad  {\crr \r\leq 0}, \label{rotations-breakdown} \\
\2 \2\pi_{\r}\circ \alpha\circ \alpha_{\mcR}\simeq \pi_{\r}\circ \alpha\,\quad \textrm{for} \quad {\crr \r>0}, \label{rotations-implementation}
\eeqa
where $\alpha_{\mcR}$, $\mcR\in SO(3)$, are the usual rotation automorphisms of scalar free field theory. Again, the extreme cases $\r\to \pm\infty$
can be extracted from the literature \cite{Ro70,KPR77}. For the interpolating family of states $\om_{\r}\circ\alpha$ we establish (\ref{rotations-implementation}) {\red and} (\ref{rotations-breakdown}) in Sections~\ref{Existence-S-matrix} {\red and} \ref{desintegration-section}, respectively.  
We conclude that the rotational symmetry, unitarily implemented in the disordered phase,  is broken  in the ordered phase, similar to the familiar case of the ferromagnetic phase transition {\red in the $O(3)$ Heisenberg model}.

\subsection{Correlation length}

Apart from constructing the Bloch-Nordsieck $S$-matrix $S_{\r}$ in the disordered phase, 
we  also obtain detailed information about the behaviour of its matrix elements near the critical point. Specifically, we obtain 
\begin{eqnarray}\label{distinction-S}
\lan \Om, S_{\r}\Om \ran\sim \left\{ \begin{array}{ll}
0 & \textrm{for\quad\quad \ \!\!\! {\crr $\r\leq  0$},}\\
\e^{-\fr{c}{\r^{3/2}} } & \textrm{\crr for $ 0 < \r\leq  1$.}
\end{array} \right.
\end{eqnarray}
To explain the  first line of (\ref{distinction-S}) let us first consider the vacuum case $\r={\crr -\infty}$. Here the standard computation with an infrared
cut-off $\si>0$ gives
\beqa
\lan \Om, S_{{\crr\r=-\infty}}^{\si}\Om \ran\sim \fr{\si}{\La},  \label{vacuum-relation}
\eeqa   
where $\La$ is a fixed UV scale introduced by $j_0$. This expression  vanishes as $\si\to 0$ and the same is true in the ordered phase {\crr ($\r<0$)}, as we show in Theorem \ref{IV-no} below.
 The second line of (\ref{distinction-S}), which we verify in Section~\ref{S-disintegration}, means, more precisely, that $C_1\e^{-\fr{c_1}{\r^{3/2}} }\leq   \lan \Om, S_{\r}\Om\ran \leq C_2\e^{-\fr{c_2}{\r^{3/2}} } $, for suitable positive constants independent of {\crr $0<\r\leq 1$}. Consequently $\lim_{\r\to {\crr 0^+}} \lan \Om, S_{\r}\Om \ran=0$, thus the dependence $\real\ni r\mapsto \lan \Om, S_{\r}\Om \ran$ is continuous, unlike the behaviour of the order parameter fluctuations in (\ref{distinction}). Relation~(\ref{vacuum-relation}) suggests {\red an interpretation of} ${\red \xi:=} \lan \Om, S_{\r}\Om \ran^{-1} $ as the correlation length of our phase transition: All photons of larger wavelengths, emitted by the external source $j$, are effectively screened by the background  radiation. This definition is adapted to our particular situation, in which even in the disordered phase the theory is genuinely massless and exponential decay 
 of correlations is not expected.  We observe from (\ref{distinction-S}), that our correlation length has a similar dependence on the control parameter $\r$  as in the case of the Berezinskii--Kosterlitz--Thouless transition.

\subsection{Phase transition  in the future lightcone} \label{lightcone-subsection}

Even if $\{\om_{\r}\}_{r\in \real}$ tend{\red s} to the vacuum as {\crr $\r\to -\infty$}, one can ask how exotic these states
are for small {\crr negative $\r$}, close to the phase transition under study. As emphasized in \cite{BR14},  the observers  have
access only to their future lightcones, so it suffices to consider this question only inside  the open future
lightcone $V_+$ with  tip at zero. Let us denote by $\mfa(V_+)$ the $C^*$-algebra generated by operators $\exp(\i \phi(f))$, where
$f \in D(V_+; \real)$. As we see from Fig.~1~(a), in the ordered phase {\crr ($\r<0$)} the 
radiation does not enter the future lightcone.  It is therefore not a surprise that in this case
the representation $\pi_{\r}$ is quasi-equivalent to the vacuum representation $\pi_{\mrm{vac}}$, i.e,
unitarily equivalent up to multiplicity \cite[Section 2.4.4.]{BR}. In a shorthand notation, we prove that
\beqa
\pi_{\r}|_{\mfa(V_+)} \quasi \pi_{\vac}  |_{\mfa(V_+)} \quad \textrm{for}\quad  \r<0.  \label{lightcone-norm}
\eeqa
This property of \emph{lightcone normality}  means physically that the observer cannot decide by experiments inside the lightcone if the radiation is in a state from  the familiar vacuum sector,
or in the state $\om_{\r}$, on a verge of the phase transition. {\red In fact, by Definition~\ref{quasiequivalent-def} below, the observer can describe the radiation by
a density matrix from the vacuum sector.}

As shown in \cite{CD20},  the relevant automorphisms $\alpha$ satisfy
\beqa
\pi_{\vac}\circ \alpha |_{\mfa(V_+)} \simeq \pi_{\vac}  |_{\mfa(V_+)}.
\eeqa   
By combining this with (\ref{lightcone-norm}), we note that these automorphisms are unitarily implemented on $\mfa(V_+)$ also in the ordered phase, perhaps
up to multiplicity. Thus the phenomenon of symmetry breaking, visible on the full Minkowski spacetime, essentially disappears in the future lightcone.
So one may ask, if the phase transition can actually be observed in $V_+$. The answer is yes, because for $f$ supported in $V_+$  the order parameters $\phi_R(f)$
are measurable in $V_+$ for all $R>0$. The distinction (\ref{distinction}) remains valid, and actually  implies
\beqa
\pi_{\r}|_{\mfa(V_+)} \not\quasi \pi_{\vac}  |_{\mfa(V_+)} \quad\textrm{for}\quad  {\crr \r\geq 0} \label{no-lightcone-norm}
\eeqa 
as we  show in Section~\ref{Absence-lightcone-normality}.   Thus by (\ref{lightcone-norm}), (\ref{no-lightcone-norm})  the property of lightcone normality is another feature that distinguishes the two phases.
Given  Fig.~1 (b) one may be surprised that lightcone normality fails at $r=0$.  In this schematic figure
the radiation {\red seems to avoid} the future lightcone also {\red at the critical point}. The reason for the failure is
that the flashes of radiation only approximately fit into the double cones at the moment of emission.  At the
critical point their tails matter {\red as we explain in Section~\ref{Lightcone-section} below estimate (\ref{two-differences-zero})}.

\subsection{Organization of the paper}

\begin{table}[t]
\centering
\begin{tabular}{|c|c|c|c|}
\hline
                 & Ordered phase $(r<0)$ &  Critical point $(r=0)$  &   Disordered phase $r>0$  \\
\hline
$S$-matrix & {\rrtable \XSolidBrush} &{\rrtable \XSolidBrush}  & {\gtable \checkmark} \\
\hline
Lightcone normality & {\gtable \checkmark} & {\rrtable \XSolidBrush} & {\rrtable \XSolidBrush} \\
\hline
\end{tabular}
\caption{Summary of main results.}
\label{table:1}
\end{table}

Our paper is organized as follows: In Section~\ref{KPR-section} we define the states $\om_r$ and reformulate them
in terms of certain symplectic maps. In Section~\ref{Existence-S-matrix} we recall the KPR mechanism for stabilizing the Bloch-Nordsieck $S$-matrix and 
show that it is at work in the entire disordered phase ($\r>0$).  Section~\ref{desintegration-section} confirms the absence of the $S$-matrix in
the ordered phase ($\r<0$) and at the critical point ($\r=0$). In Section~\ref{S-disintegration} we have a closer look at the disintegration of the
$S$-matrix near the critical point and derive formula~(\ref{distinction-S}) for the correlation length. In Section~\ref{Absence-lightcone-normality} we verify the infinite fluctuations
of the timelike order parameter in the disordered phase and at the critical point. We  conclude  the  absence of lightcone normality. We also
demonstrate infinite fluctuations of the spacelike order parameter in the disordered phase and at the critical point. {\red We  point out that infinite fluctuations
of the spacelike order parameter are a necessary condition for the existence of  the $S$-matrix.} 
Section~\ref{Lightcone-section}  is devoted to  lightcone normality in the ordered phase, which is our main technical result. In Section~\ref{ordered-section} we indicate that lightcone normality
implies vacuum fluctuations of the timelike order parameter, as stated in the first line of (\ref{distinction}). More surprisingly, certain technical ingredients
from the proof of lightcone normality  also ensure vacuum fluctuations of the spacelike order parameter.  Recalling from above that finite fluctuations of the spacelike
order parameter are in conflict with the KPR mechanism, this sheds some light on the intriguing incompatibility between the existence of the Bloch-Nordsieck $S$-matrix
 and lightcone normality, see Table~\ref{table:1}.

\vspace{0.2cm}

\noindent{\bf {\rtable Acknowledgement}:} We would like to thank Detlev Buchholz for useful discussions and pointing out the reference \cite{DFG84}. We also thank
 Daniela Cadamuro and Henning Bostelmann for helpful discussions.  
 Financial support from the  grant  `Preludium' 2021/41/N/ST1/02755  of the National Science Centre, Poland, is gratefully acknowledged.

\section{Interpolating states $\om_{\r}$} \label{KPR-section}
\setcounter{equation}{0}

In this section we give a definition  of the states $\om_{\r}$ which interpolate between the vacuum at {\crr $\r=-\infty$} and the KPR infravacuum at {\crr $\r=\infty$}. We start from the following family of 
modes of radiation, introduced first in \cite{Re74,KPR77}:
\begin{itemize}

\item We introduce  sequences $\eps_i:=2^{-(i-1)}$ and $b_i:= \fr{1}{i}$ for $i=1,2,3\ldots$

\item We define normalized functions $\xii_i(|\vk|):=c_0\fr{\chi_{[\eps_{i+1},\eps_i] }(|\vk|)}{|\vk|^{3/2}}\in L^2(\real_+, |\k|^2d|\k|)$, where $\chi_{[\eps_{i+1},\eps_i]}$ is the
characteristic function of $[\eps_{i+1},\eps_i]$  and  $c_{0}:=(\log 2)^{-1/2}$
 is the normalization constant.

\item We define the sets
\beqa
I_n:=\{\, \al=(i,\ell,m) \,|\, i={\red 1},2,3,4\ldots n, \ 0\leq \ell \leq i, \ -\ell\leq m\leq \ell \ \} \label{set-I-n}
\eeqa 
and let $\zeta_{\al}(\k):=\xii_i(|\k|) Y_{\ell m}(\hat \k)$ for $\al\in I_n$,   where $Y_{\ell m}$ are the real valued spherical harmonics.

\end{itemize}
As explained in Subsection~\ref{description-transition}, we want to create flashes of radiation at times $-\taur_{\al}:=-\taur_i$
defined in~(\ref{taur-i}). Thus we introduce the time-translated KPR-modes
\beqa
\hzeta_{\al}(\k):=\e^{-\i  |\k|\taur_{\al} }\zeta_{\al}(\k).
\eeqa
A family of $n$ flashes of radiation, as considered in Subsection~\ref{description-transition}, is described by the following squeezed state of radiation
\beqa
\Psir_n=\prod_{\al\in I_n} \exp\bigg( -\fr{1}{\ww_{\al}} a^*( \hzeta_{\al})^2   \bigg)\Om, \label{psirn}
\eeqa
 where $\ww_{\al}:=\fr{b_i^2+1}{b_i^2-1}$ and we omitted normalization\footnote{\red For $i=1$ we have $\ww_{\al}=\infty$ thus  $1/ \ww_{\al}=0$ in (\ref{psirn}). So we could have omitted $i=1$ mode from the beginning. Actually, omitting any finite number of modes does not change our conclusions, cf. Appendix~\ref{omitting}.}. Now the states $\om_{\r}$, consisting of infinitely many flashes,  are given by
\beqa
\om_{\r}(A)=\lim_{n\to \infty} \fr{\lan \Psir_n, A  \Psir_n\ran}{\lan  \Psir_n,  \Psir_n\ran}, \quad A\in \mfa. \label{def-om-r}
\eeqa

The above representation of the  states $\om_{\r}$ is closest in spirit to the discussion from Subsection~\ref{description-transition},
but is not always convenient for computations. Therefore, we rearrange it as follows: Let us note that by Gaussian integration
\beqa
\fr{\Psir_n }{\|\Psir_n\| }=  \mcUr(I_n)\Om  :=\prod_{\al \in I_n}\bigg( \fr{ z_{\al} }{2\pi } \bigg)^{1/2} 
\int ds_{\al} \, \e^{- \hh (\ww_{\al}  -1)s_{\al}^2} W(\i s_{\al}\hzeta_{\al})\Om, \label{Weyl-operator-implementation}
\eeqa 
where $z_{\al}:=\h\sqrt{\ww_{\al}^2-1}$ and $W(\zeta):=\exp(\i \fr{1}{\sqrt{2}}(a^*(\zeta) +a(\zeta) )  )$, $\zeta\in L^2(\real^3)$, {\red are} the Weyl operators. Inserting this identity 
to (\ref{def-om-r}), using the basic relations 
\beqa
W(\zeta) W(\zeta')=\e^{-\h \i \mrm{Im} \lan \zeta,\zeta'\ran } W(\zeta+\zeta'), \quad \lan \Om, W(\zeta)\Om\ran=\e^{-\fr{1}{4}\|\zeta\|_2^2}, \label{Weyl+vacuum}
\eeqa
and performing the resulting Gaussian integrals, we conclude that
\beqa
\om_{\r}(W(F))=\lan \Om, W( \Tr F)\Om\ran. \label{om-r-def-one}
\eeqa
Here $\Tr$ is a certain symplectic map on the  space $\mcL:=\{\,F=\phi(f)\Om\,|\, f\in D(\real^4;\real)\}$ equipped with the
symplectic form  $\mrm{Im}\lan\,\cdot\,, \cdot\,\ran$.  Then the GNS representation of $\om_{\r}$ acts naturally on $\mcF$ by
\beqa
\pi_{\r}(W(F))=W(\Tr F). \label{om-r-def-two}
\eeqa 
The symplectic map has the form
\beqa
\Tr= 1+   \sum_{i=1}^{\infty}(b_{i}^{-1}-1)  u_{-\taur_i} Q_i \mrm{Re} \,u_{\taur_i}+ \i \sum_{i=1}^{\infty} (b_{i}-1) u_{-\taur_i} Q_i \mrm{Im}\,u_{\taur_i}, \label{T-r-def}
\eeqa
where $u_{\tau}:=\e^{\i \mu \tau}$, with $\mu(\k):=|\k|$, are time translations and $Q_i$ are orthogonal projections given by
\begin{equation}\label{pmbQi}
Q_i = |\xii_i\ran\lan  \xii_i| \otimes \tilde{Q}_i \quad \text{with} \quad \tilde{Q}_i := 
\sum_{0\leq  \ell \leq i} 
\sum_{m=-\ell}^\ell   | Y_{\ell m } \rangle \langle Y_{\ell m } |.
\end{equation}
We note in passing that above reasoning is an instance of the general theory of implementation of symplectic maps \cite{Sh62, SS65, Ru78, AY82}.
In the representation (\ref{om-r-def-one}), (\ref{T-r-def}) it is easy to see that $\lim_{r\to-\infty} \om_r(W(F))=\om_{\mrm{vac}}(W(F))$. In fact, due to 
large oscillations only $1$ from (\ref{T-r-def}) survives  this limit as a consequence of the  Riemann-Lebesgue lemma and dominated convergence. 
It is also clear that the limit $r\to \infty$ amounts to replacing $u_{\pm \taur_i}$ with $1$ in (\ref{T-r-def}).   Then we recover the usual KPR maps $\T$ from \cite{KPR77}.

From formulas~(\ref{om-r-def-one})--(\ref{pmbQi}) it is also manifest that  states $\om_{\r}$ are invariant under rotations. Let $(u_{\mcR} F)(\k):=F_{\mcR}(\k):= F(\mcR^{-1}\k)$, $\mcR\in SO(3)$,
be the standard  representation of rotations on $L^2(\real^3)$. It gives rise to automorphisms of $\mfa$ via $\alpha_{\mcR}(W(F))=W(F_{\mcR})$. Since $\ti Q_i$ are spectral
projections of the total angular momentum operator $\mrm{L}^2$, we easily see that $\Tr \circ u_{\mcR}=u_{\mcR}\circ \Tr$. This immediately gives  $\om_{\r}\circ \alpha_{\mcR}=\om_{\r}$, hence
\beqa
\pi_{\r}\circ \alpha_{\mcR}(\,\cdot\,)=U_{\mcR}\pi_{\r}(\,\cdot\, ) U_{\mcR}^* \label{rotations-implementation-x}
\eeqa
for a unitary representation of rotations $\mcR\mapsto U_{\mcR}$ on $\mcF$.

\section{The existence of the $S$-matrix in the disordered phase} \label{Existence-S-matrix}
\setcounter{equation}{0}

The main technical result of this section is the following:
\bet\label{IV-yes}    $\Tr \vv \in L^2(\real^3)$  for all $\vv$ as in (\ref{vv-def}) and  {\crr $\r>0$}.
\eet
\nin Here we wrote $\vv:=\vv^{\mrm{out}}$ for brevity, and to stress that the general form of these functions is the same in the `in' and `out' case. 
Since $\vv$ is not in $\mcL$ and not even in $L^2(\real^3)$ due to a singularity at $\k=0$, the operation $\Tr \vv$ requires clarification:
We introduce an intermediate infrared cut-off $\si>0$ and set $\vv_{\si}(\k):=\chi_{[\si,\infty)}(|\k|)\vv(\k)$, where  $\chi$ is the characteristic function. 
Then, more precisely, Theorem~\ref{IV-yes} requires that the limit $\Tr \vv:=\lim_{\si\to 0} \Tr\, \vv_{\si}$ exists in $L^2(\real^3)$.

Before we prove Theorem~\ref{IV-yes}, let us indicate its relevance to the problem of the existence of the Bloch-Nordsieck $S$-matrix discussed in Subsection~\ref{spontaneous-symmetry-breaking}. By considering relation (\ref{alpha-automorphisms}) in representation $\pi_{\r}$ we obtain
\beqa
\pi_{\r}\circ \alpha(W(F) )\2=\2 \e^{-\i \mrm{Im}\lan \vv^{\mrm{out}}-  \vv^{\mrm{in}}, F\ran }\pi_{\r}(W(F))\non\\
\2=\2  \e^{-\i \mrm{Im}\lan \Tr (\vv^{\mrm{out}}-  \vv^{\mrm{in}}), \Tr F\ran }\pi_{\r}(W(F)) \non\\
\2=\2 W(\Tr (\vv^{\mrm{out}}-  \vv^{\mrm{in}}) ) \pi_{\r}(W(F))W(\Tr (\vv^{\mrm{out}}-  \vv^{\mrm{in}}) )^*, \label{S-matrix-computation}
\eeqa
where in the second step we used that $\Tr$ is symplectic and in the last step we exploited the Weyl relations (\ref{Weyl+vacuum}). Thus, by Theorem~\ref{IV-yes}, 
the $S$-matrix $S_{\r}:=W(\Tr (\vv^{\mrm{out}}-  \vv^{\mrm{in}}) )$ is well defined. This is how the $S$-matrix is stabilized by the KPR mechanism.
Combining (\ref{S-matrix-computation}) with (\ref{rotations-implementation-x}) we also obtain
\beqa
\pi_{\r}\circ \alpha \circ \alpha_{\mcR}(W(F))=S_{\r}U_{\mcR} S_{\r}^*\, \pi_{\r}\circ \alpha (W(F))\, S_{\r}U_{\mcR}^* S_{\r}^*,
\eeqa
which confirms (\ref{rotations-implementation}).

The first step of the proof of Theorem~\ref{IV-yes} is to check the case {\crr $\r=\infty$}. Then the time translations disappear in (\ref{T-r-def}) and the  
maps $\Tr$ become simply the KPR maps $\T$. Noting that $\vv$ is purely imaginary, we have
\beqa
\T \vv= (1+\sum_{i=1}^{\infty} (b_{i}-1) Q_i) \vv. \label{KPR-relation}
\eeqa
The somewhat intricate definition of the KPR states and maps from \cite{KPR77} is motivated precisely by the requirement that the above
vector is in $L^2(\real^3)$. As there is a detailed modern account in \cite[Proposition 4.2]{CD18}, we can be brief here. We can replace $\vv$  from (\ref{vv-def})
with
\beqa
\vv'(\k):= \i q \chi_{[0,1]}(|\k|) \,  \fr{1}{ |\vk|^{3/2}(1-\hat{\k} \cdot \v ) }, \label{chi-variant}
\eeqa
where $q$ appeared below (\ref{j-def}), since the difference is an infrared regular vector from $L^2(\real^3)$.  We have
\beqa
\T \vv'=\sum_{i=1}^{\infty} b_i Q_i\vv'+(\vv'-\sum_{i=1}^{\infty} Q_i\vv').
\eeqa
One can show that the first term on the r.h.s. is in $L^2(\real^3)$ using the fact that $b_i\to 0$. 
As for the second term, we exploit that $\vv'=\sum_{i=1}^{\infty}(|\xii_i\ran \lan \xii_i|\otimes 1)\vv'$ {\red thanks to $\chi_{[0,1]}$ in  (\ref{chi-variant})}. Thus we obtain 
\beqa
(\vv'-\sum_{i=1}^{\infty} Q_i\vv')=\sum_{i=1}^{\infty}\sum_{\ell > i} \sum_{m=-\ell}^\ell  (|\xii_i\ran \lan \xii_i|\otimes  
  | Y_{\ell m } \rangle \langle Y_{\ell m } |    )  \vv'. \label{KPR-sum}
\eeqa  
Now using $\lan Y_{\ell m }, \vv'\ran=\fr{1}{ \sqrt{\ell(\ell+1)}} \lan   Y_{\ell m },  \mrm{L}^2\vv'\ran$ and $\ell>i$ we {\red get}
sufficiently strong convergence of the sum to conclude that  (\ref{KPR-sum}) is also in $L^2(\real^3)$. This gives Theorem~\ref{IV-yes} for $r=\infty$.

In order to prove Theorem~\ref{IV-yes} for arbitrary {\crr $\r>0$}, we {\red come back to definition} (\ref{T-r-def}) of $\Tr$. Recalling that  $\vv$ as in (\ref{vv-def}) 
is purely imaginary, we have
\beqa
\Tr \vv 
\2=\2 \vv+   \sum_{i=1}^{\infty}(b_{i}^{-1}-1)  u_{-\taur_i} Q_i ( \i\,\sin(\mu \taur_i)  \vv)+ \i \sum_{i=1}^{\infty} (b_{i}-1) u_{-\taur_i} Q_i 
(-\i\cos( \mu \taur_i) \vv). \label{T-v-simple}
\eeqa
Exploiting (\ref{KPR-relation}), we can write
\beqa
\Tr \vv\2=\2(\Tr- \T)\vv+\T\vv \non\\
\2=\2 \sum_{i=1}^{\infty}(b_{i}^{-1}-1)  u_{-\taur_i} Q_i ( \i\,\sin(\mu \taur_i)  \vv)+ \sum_{i=1}^{\infty} (b_{i}-1)
 [u_{-\taur_i}Q_i (\cos( \mu \taur_i)-1)\vv ] \label{two-sums}\\
\2  \2\phantom{444444444444444444444444444444}{\red +}\sum_{i=1}^{\infty}(b_{i}-1)  (u_{-\taur_i}-1) Q_i  \vv+\T\vv. \label{one-sum}
\eeqa
We know already that $\T\vv\in L^2(\real^3)$, so it suffices to consider the remaining terms. In the case of (\ref{two-sums})
it is clear that $\sin(\mu \taur_i)\sim \mu\taur_i $ and $\cos( \mu \taur_i)-1\sim (\mu \taur_i)^2$ have a regularizing effect
on the $\mu^{-3/2}$ singularity of $\vv$. We note the estimates {\red on the $L^2$-norms} for $l=1,2$ 
\beqa
\|Q_i (\mu \taur_i)^{l}\vv\|_2\leq (\taur_i)^l \|\chi_{[\eps_{i+1},\eps_i]}(\mu) \mu^l \vv\|_2\leq c (\taur_i \eps_i)^l=c2^{ {\crr -\r(i-1)l}} \label{l-bound}
\eeqa
for a constant $c$ independent of $i$. Using this, we immediately obtain the convergence in $L^2(\real^2)$ of the sums in (\ref{two-sums}) {\crr for $\r>0$}.

The regularizing effect of $ (u_{-\taur_i}-1)\sim \mu \taur_i$ in (\ref{one-sum})  is slightly less obvious as this operator does not act directly on $\vv$.
We argue as follows: On the one hand,
\beqa
 \|(u_{-\taur_i}-1) Q_i\|\leq \taur_{i}\|\mu Q_i\|\leq \taur_{i}\|\mu \xii_i\|_2\leq c\taur_{i}\eps_i=c2^{{\crr -\r(i-1)}}, \label{one-hand}
 \eeqa
{\red where $\|\,\cdot\,\|$ above is the operator norm.} On the other hand
\beqa
\|Q_i \vv\|_2\leq \|\chi_{[\eps_{i+1},\eps_i]}(\mu) \vv\|_2\leq c.   \label{other-hand}
\eeqa
Using (\ref{one-hand}) and (\ref{other-hand}),  the $L^2$-convergence of 
the sum in (\ref{one-sum})  follows {\crr for $\r>0$}. This completes the proof of Theorem~\ref{IV-yes}.

\section{Non-existence of the $S$-matrix  in the ordered phase  and at the critical point} \label{desintegration-section}
\setcounter{equation}{0}

The main technical result of this section is the following theorem, which relies on the assumption $0\neq q:=\int d^3\x\, j_0(\x)$. 
\bet\label{IV-no} {\crr Let $\r\leq  0$}. Then    $ \lim_{\si\to 0}\|\Tr (\vv^{\mrm{out}}_{\si}-\vv^{\mrm{in}}_{\si})\|_2=\infty$  for all $\vv^{\mrm{in}/\mrm{out}}$ as in (\ref{vv-def}) and   $\v^{\mrm{in}}\neq \v^{\mrm{out}}$.
\eet

Let us first indicate, following \cite{Ro70}, how the property from Theorem~\ref{IV-no} prevents the existence of the $S$-matrix.  
Coming back to computation (\ref{S-matrix-computation}), suppose by contradiction that there exists  a unitary operator $S_{\r}$ such that
\beqa
 \e^{-\i \mrm{Im}\lan \Tr (\vv^{\mrm{out}}-  \vv^{\mrm{in}}), \Tr F\ran }\pi_{\r}(W(F)) =  S_{\r}  \pi_{\r}(W(F)) S_{\r}^*. \label{S-matrix-computation-x}
\eeqa
By evaluating both sides on $\lan \Om, \, \cdot \,\, \Om\ran$ and using (\ref{Weyl+vacuum}), (\ref{om-r-def-two}), we obtain
\beqa
\e^{-\i \mrm{Im}\lan \Tr (\vv^{\mrm{out}}-  \vv^{\mrm{in}}), \Tr F\ran }=\e^{\fr{1}{4}\|\Tr F\|_2} \lan \Om, S_{\r}  W(\Tr F) S_{\r}^*\Om\ran.
\eeqa
Treating both sides as functions of $\Tr F$ we note that the l.h.s. is discontinuous in the topology of $L^2(\real^3)$ by Theorem~\ref{IV-no} 
and the Schur lemma. However, the r.h.s. is continuous by the continuity of the maps $L^2(\real^3)\ni \zeta\mapsto W(\zeta)\Psi$, $\Psi\in \mcF$, between Hilbert spaces. Using that $\Tr: \mcL\to L^2(\real^3)$ has a dense range  (cf. \cite[Lemma 3.3]{Ku98})  we obtain a contradiction.  

By a similar token we obtain the rotational symmetry breaking in the ordered phase. Let $\alpha_{\v}$ be the automorphism {\red $\alpha$} corresponding to $\vv$ given by (\ref{vv-def})
with $\v:=\v^{\mrm{out}}$. Let us  assume for simplicity that $\v^{\mrm{in}}=0$. Now suppose that there is a unitary $V_{\mcR}$ such that
\beqa
\pi_{\r}\circ \alpha_{\v}\circ \alpha_{0}^{-1}\circ\alpha_{\mcR}({\red A})=V_{\mcR}\pi_{\r}\circ \alpha_{\v}\circ \alpha_{0}^{-1}({\red A}) V_{\mcR}^*, \quad {\red A\in \mfa}.
\eeqa 
Then, by exploiting (\ref{rotations-implementation-x}), noting $\alpha_{\mcR^{-1}}\circ \alpha_{\v} \circ \alpha_{\mcR}=\alpha_{\mcR \v}$ {\red and redefining 
$\alpha_{0}^{-1}(A)\to A$, we obtain}
\beqa
\pi_{\r}\circ \alpha_{\mcR \v} \circ\alpha_{\v}^{-1}({\red A})=U_{\mcR}^*V_{\mcR}\pi_{\r}({\red A}) V_{\mcR}^*U_{\mcR}, \quad {\red A\in \mfa}.
\eeqa
By evaluating this relation  on a Weyl operator we obtain an identity which is analogous to (\ref{S-matrix-computation-x}). Thus we arrive at
a contradiction which confirms (\ref{rotations-breakdown}).

Let us now move on to the proof of Theorem~\ref{IV-no}. As in the proof of Theorem~\ref{IV-yes}, we can replace $\vv^{\mrm{in}/\mrm{out}}$ with 
$\vv'{}^{\mrm{in}/\mrm{out}}$ which have the factor $\ti j_0(\k)$ replaced with $q\chi_{[0,1]}(\k)$ as the differences are infrared regular vectors in $L^2(\real^3)$.  
We set  $\vv=\vv'{}^{\mrm{out}}-  \vv'{}^{\mrm{in}}$ and compute $\|\Tr \vv_{\si}\|_2$  using (\ref{T-v-simple})
\beqa
\|\Tr \vv_{\si}\|_2=\bigg\|\sum_{i=1}^{\infty} u_{-\taur_i} \bigg\{ u_{\taur_i}\chi_{\De_i} \vv_{\si}+  (b_{i}^{-1}-1)   Q_i ( \i\,\sin(\mu \taur_i)  \vv_{\si})+ \i  (b_{i}-1)  Q_i 
(-\i\cos( \mu \taur_i) \vv_{\si})\bigg\}\bigg\|_2,\,\, \label{reasoning-beginning}
\eeqa
where $\chi_{\De_i}$ are characteristic functions of the concentric shells $\De_i:=\{ \k\in \real^3\,|\, \eps_{i+1}\leq |\k|\leq \eps_i\, \}$.
Since the vectors in curly brackets are supported in distinct shells, we can drop the action of $u_{-\taur_i}$ under the norm. Decomposing
these vectors into real and imaginary parts, we obtain a lower bound
\beqa
\|\Tr \vv_{\si}\|_2 \2\geq\2 \bigg\|\sum_{i=1}^{\infty} \bigg\{  \i \sin(\mu \taur_i) \chi_{\De_i} \vv_{\si}+  (b_{i}^{-1}-1)   Q_i ( \i\,\sin(\mu \taur_i)  \vv_{\si})\bigg\}\bigg\|_2.
\eeqa
Hence, noting that the cut-off $\si$ truncates the sum to some finite $n(\si)$, such that $\lim_{\si\to 0} n(\si)\to \infty$, we have 
\beqa
\|\Tr \vv_{\si}\|^2_2\2\geq \2 \sum_{i=1}^{n(\si)}  \lan \vv_{\si},  \chi_{\De_i}\sin^2(\mu\taur_i) \vv_{\si} \ran   +\sum_{i=1}^{n(\si)}(b_i^{-2}-1) \lan \vv_{\si},  \sin(\mu\taur_i)Q_i \sin(\mu\taur_i) \vv_{\si}\ran.  \label{sum-blows-up}
\eeqa
Since both sums above are positive, it suffices to show that the first of them blows up in the limit $\si\to 0$. {\crr For $\r<0$} we have 
\beqa
\lan \vv_{\si},  \chi_{\De_i}\sin^2(\mu\taur_i) \vv_{\si} \ran\2=\2 c\int_{\eps_{i+1}}^{\eps_i} \fr{d|\k|}{|\k|}  \sin^2( 2^{ {\crr - \r}(i-1)} |\k|/\eps_{i} )\non\\
\2=\2c \int_{1/2}^{1} \fr{d|\k|}{|\k|}  \sin^2(2^{ {\crr -\r}(i-1)} |\k|)\underset{i\to \infty}{\to} c\int_{1/2}^{1} \fr{d|\k|}{2|\k|}. \label{no-S-matrix}
\eeqa
Here $c>0$ is an inessential factor coming from angular integration and the last step follows from
$\sin^2(x)=\fr{1}{2}(1-\cos(2x))$ and the Riemann-Lebesgue lemma. For $\r=0$ the expression on the l.h.s. of  (\ref{no-S-matrix}) is independent of $i$ and non-zero. Thus the sum in (\ref{sum-blows-up}) diverges  {\red as $\si\to 0$} for $r\leq 0$. This concludes the proof of Theorem~\ref{IV-no}.

\section{The $S$-matrix near the critical point}\label{S-disintegration}
\setcounter{equation}{0}

Recall that $S_{\r}$ denotes the Bloch-Norsieck $S$-matrix in the disordered phase $({\crr \r>0})$.
In this section we describe the disintegration of this $S$-matrix  as we approach the critical
point, that is as $\r\to {\crr 0^+}$. Specifically, we will establish the second line in (\ref{distinction-S}).

For this purpose we obtain {\red from  {\red (\ref{S-matrix-computation}), (\ref{Weyl+vacuum})} } 
\beqa
\lan \Om, S_{\r}\Om\ran=\e^{-\fr{1}{4}\|\Tr \vv\|^2_2}
\eeqa
and  recall that the quantity $\|\Tr \vv\|^2_2$ was the subject of our interest in Sections~\ref{Existence-S-matrix}, \ref{desintegration-section}. {\red (We allow a modification
of $g$ as in (\ref{chi-variant}), as it cannot affect the singular behaviour of $S_r$ near the critical point).}
In particular, {\red formula (\ref{two-sums})--(\ref{one-sum}) and} estimates (\ref{l-bound})-(\ref{other-hand}) immediately give the following upper bound for {\crr  $1\geq \r>0$}
\beqa
\|\Tr \vv\|_2\2{\red\leq}\2 { \| \sum_{i=1}^{\infty}(b_{i}^{-1}-1)  Q_i ( \i\,\sin(\mu \taur_i)  \vv)\|_2+ \sum_{i=1}^{\infty} |b_{i}-1|
 \|Q_i (\cos( \mu \taur_i)-1)\vv\|_2 } \label{two-sums-x}\\
\2  \2\phantom{4444444444444444444444444444}{ {\red+}\sum_{i=1}^{\infty}|b_{i}-1| \| (u_{-\taur_i}-1) Q_i  \vv\|_2+\|\T\vv\|_2} \non\\
\2 \leq \2\bigg( c\sum_{i=1}^{\infty}(b_{i}^{-1}-1)^2  2^{{\crr -2\r} (i-1)} \bigg)^{1/2}   +c\sum_{i=1}^{\infty} 2^{ {\crr -2\r} (i-1)}+c\sum_{i=1}^{\infty} 2^{{\crr -\r}(i-1)}+c  \leq 
\fr{c'}{\r^{3/2}}, \label{corr-length-dep} 
\eeqa
{\red where the constants $c,c'$ are independent of $r$.}

Now we are looking for a lower bound for this quantity. We observe, that in the reasoning (\ref{reasoning-beginning})--(\ref{no-S-matrix})
we used the assumption $\r \leq 0$ only in the very last step, while the preceding steps hold for arbitrary $\r\in \real$. As the first sum on the r.h.s. of
(\ref{sum-blows-up}) leads to a weak upper bound, which does not match the dependance (\ref{corr-length-dep}), we look at the second
sum in (\ref{sum-blows-up})
\beqa
\|\Tr \vv \|^2_2\2\geq \2 \sum_{i=1}^{\infty}(b_i^{-2}-1) \lan \vv,  \sin(\mu\taur_i)Q_i \sin(\mu\taur_i) \vv\ran.  \label{lower-bound-preparation}
\eeqa
We used here the fact that we are in the disordered phase and the bounds  (\ref{l-bound})-(\ref{other-hand}) {\red allow us} to take the limit $\si\to 0$.
Next, we compute
\beqa
\lan \vv,  \sin(\mu\taur_i)Q_i \sin(\mu\taur_i) \vv\ran 
\2\geq \2 c_i (2^{ {\crr -2\r}})^{i-1} \bigg| \int_{\h }^{1}  d|\k| \fr{ \sin(|\k| (2^{{\crr-\r}})^{i-1}) }{ |\k| (2^{{\crr-\r}})^{i-1}} \bigg|^2\non\\
\2 \geq \2 c_i (2^{{\crr -2\r}})^{i-1} \fr{1}{16},
\eeqa 
where in the last step we used $\fr{\sin x}{x}\geq 1/2$ for $|x|\leq 1$.
The constants $c_i$ arise from angular integration. As they have the form $\lan \vv_{\mrm{ang}}, \ti Q_i \vv_{\mrm{ang}}\ran$, where
$\vv_{\mrm{ang}}\neq 0$ is the angular part of $\vv$, and $\ti Q_i \nearrow 1_{L^2(S^2)}$ is an increasing family of projections, we know
that $c_i\geq c_{i_0}>0$  for some finite $i_0$. On the other hand, first few $c_i$ may vanish for certain choices of $\v^{\mrm{in}} \neq \v^{\mrm{out}}$.
Coming back to (\ref{lower-bound-preparation}), we can  write
\beqa
\|\Tr \vv \|^2_2\2\geq \2  \fr{c_{i_0}}{16}\sum_{i=i_0}^{\infty}(i^2-1) (2^{{\crr -2\r}})^{i-1}
\geq \fr{c_{i_0}}{16}  \fr{2\e^{-2\xn}}{(1-\e^{-\xn})^3}  \e^{-\pn \xn} |_{a=2|\r|\log(2)} \geq \fr{c}{\r^3}.
\eeqa
Thus we have verified (\ref{distinction-S}).

\section{The absence of lightcone normality in the disordered phase  and at the critical point} \label{Absence-lightcone-normality}
\setcounter{equation}{0}

The main result of this section is the following:
\bet \label{no-lightcone-normality} The representations $\pi_{\r}$ are not lightcone normal for {\crr $\r\geq 0$}. 
\eet

The breakdown of lightcone normality in the disordered phase {\red and at the critical point}, we alluded to in formula~(\ref{no-lightcone-norm}),
is driven by infinite fluctuations of the timelike order parameter $\phi_R(f)$, which we stated in  (\ref{distinction}).
For future convenience we note the relation $\e^{\i \phi_R(f) }=W(F_R)$, 
where
\beqa
F_{R}(\k):=R^{3/2} F(\k R), \quad F(\k)=(2\pi)^{-3/2} |\k|^{-1/2} \int \e^{\i |\k|t-\i \k\cdot \x } f(t,\x) dt {\rtable d^3\x} \label{central-sequence-def}
\eeqa
{\red and $f$ is supported in the future lightcone $V_+$.}  Using this notation, we can write
\beqa
\lan \Om,\pi_{\r}(W(F_R))\Om\ran=\e^{-\fr{1}{4}\|\Tr F_R\|^2_2}=\e^{-\fr{1}{4}\om_{\r}(\phi_R(f)\phi_R(f))} \label{W-F-R}
\eeqa
and the last expression tends to zero if timelike fluctuations of the order parameter tend to infinity with $R$.
By Lemma~\ref{central-lemma} below, this property prevents lightcone normality of the representation.

In Subsection~\ref{lightcone-subsection}  we defined the property of lightcone normality of a given representation $\pi$  as 
 quasi-equivalence of $\pi$ and $\pi_{\vac}$ on  $\mfa(V_+)$ {\red that is} unitary equivalence up to multiplicity. 
In this section it is  more convenient to use a different (but  equivalent) definition \cite[Section 2.4.4.]{BR}: 
\bed\label{quasiequivalent-def} If $\pi$ is a representation of a $C^*$-algebra  $\mfa$ then a state $\om$ of $\mfa$
is said to be $\pi$-normal if there exists a density matrix $\rho$   such that
\beqa
\om(A)=\mrm{Tr}(\rho \pi(A) )
\eeqa
for all $A\in \mfa$. Two representations $\pi_1$ and $\pi_2$ of $\mfa$ are said to be 
quasi-equivalent, written $\pi_1\quasi \pi_2${\rtable , if each $\pi_1$-normal state} is $\pi_2$-normal
and conversely.
\eed
The following lemma, which is an instance of the method of central sequences, 
gives a useful criterion to exclude lightcone normality. 
The irreducibility of  $\pi_{\r}$, which is {\red stated} as an assumption, is a consequence of the dense range of $\Tr$. For the latter property we refer to 
\cite[Lemma 3.3]{Ku98}.
 \bel\label{central-lemma} Let $\pi: \mfa\to B(\mcF)$ be an irreducible representation. Suppose that the following limit exists (possibly along a subsequence) and
 \beqa
 \lim_{R \to \infty} \lan \Om, \pi(W(F_{R}))\Om\ran \neq  \lan \Om, W(F)\Om\ran. \label{central-sequence-relation}
 \eeqa
  Then $\pi$ is not lightcone normal.
  \eel
 \proof Suppose, by contradiction, that $\pi$ is lightcone normal. Then $\om_{\vac}$ is $\pi$-normal,
 hence there exists a density matrix $\rho$ such that 
 \beqa
 \lan \Om, W(F)\Om\ran=\lan\Om, W(F_{R}) \Om\ran=\mrm{Tr}\big( \rho\, \pi(W(F_{R})) \big)=\sum_{j=1}^{\infty} p_j \lan \Psi_j,    \pi(W(F_{R})) \Psi_j\ran,   \label{key-relation}
 \eeqa 
 where we expressed $\rho=\sum_{j=1}^{\infty}p_j |\Psi_j\ran \lan \Psi_j|$ via an  orthonormal system $\{\Psi_j\}_{j\in \nat}$ in  $\mcF$ and  $\{p_j\}_{j\in \nat}$  
 {\red is a sequence of positive numbers, summable to one.}  Now we can write  $\Psi_j=U_j\Om$ for some unitaries $U_j\in \pi(\mfa)$ by the  irreducibility assumption and the Kadison transitivity theorem \cite[Theorem~10.2.1]{KR}.  
 We observe that the localization region of $W(F_{R})=\e^{\i \phi_R(f)} $ is shifted to  timelike infinity as $R\to \infty$. Consequently, by  the Huyghens principle, i.e.
 commutation of the massless scalar free field at timelike separation of arguments,  $\lim_{R\to \infty}[W(F_{R}), A]= 0$ in norm  for any $A\in \mfa$. This gives
 \beqa
 \lim_{R \to \infty} \lan \Psi_j,    \pi(W(F_{R})) \Psi_j\ran=\lim_{R\to \infty} \lan \Om,    \pi(W(F_{R})) \Om\ran.
\eeqa  
Now by taking the limit $R \to \infty$ in  (\ref{key-relation}) and using (\ref{central-sequence-relation}), we conclude the proof. \qed

Let us now prove Theorem~\ref{no-lightcone-normality} using Lemma~\ref{central-lemma}. Due to (\ref{W-F-R}) it suffices to show that 
$ \lim_{R\to \infty}\|\Tr F_R\|_2=\infty$ along a subsequence. Coming back to representation (\ref{T-r-def}), we can write
\beqa
\|\Tr F_R\|_2\geq \|  \sum_{i=1}^{\infty}(b_{i}^{-1}-1)  u_{-\taur_i} Q_i \mrm{Re} \,u_{\taur_i} F_R\|_{ {\red 2} }+O(1). \label{modified-discussion-start}
\eeqa
{\red Here} $O(1)$ denotes error terms which are manifestly bounded in $R$, because the respective operators are bounded and
$\|F_R\|_2$ is independent of $R$.
Thus it suffices to show that
\beqa
\lim_{n\to \infty}\sum_{i=1}^{\infty}(b_i^{-1}-1)^2 \lan \mrm{Re} (u_{\taur_i} F_{R_n}), Q_i \mrm{Re} (u_{\taur_i} F_{R_n})\ran=\infty, \label{to-verify-lightcone-norm}
\eeqa
for some subsequence $\{R_n\}_{n\in \nat}$, tending to infinity.
As we are interested in a bound from below, it suffices to consider the $n$-th term in the sum over $i$  and the $\ell=0$ term  
in the sum over the spherical harmonics hidden in $Q_i$. Let us pick in (\ref{central-sequence-def})  $f(t,\x)= \de(t-\tau) \eta(\x)$, where  
$\tau>0$ is sufficiently large depending on the support of $\eta$, so that $f$ is supported inside the future lightcone.
 Furthermore, we choose $\eta$ spherically symmetric and such that  $\ti\eta$  is positive. 
 The resulting expression reads
\beqa
  (b_n^{-1}-1)^2 |\lan \xii_n, \mrm{Re}({\red u_{\taur_i}}F_{R_n}) \ran|^2 
 = c_0^2(b_n^{-1}-1)^2\bigg( \int_{ \eps_{n+1} R_n  }^{\eps_n R_n  }  d|\ti{\k}| \,     \ti{\eta}(|\ti{\k}|)  
 \cos( |\ti{\k}| R_n^{-1} (2^{-{\red (n-1)}|r|}  \eps^{-1}_n+\tau )) \bigg)^2. \,\,
 \label{strictly-positive-integral}
  \eeqa
Now we choose $R_n:= L/\eps_n$ for some $L>0$. First, suppose that {\crr $\r>0$}  and compute
\beqa
\lim_{n\to \infty}\int_{ \eps_{n+1} R_n  }^{ \eps_nR_n }  d|\ti{\k}| \,  \ti{\eta}(\ti{\k})  \cos( |\ti{\k}| R_n^{-1} (2^{-{\red (n-1)} |r|}   \eps^{-1}_n +\tau))   =
\int_{L/2}^{ L }  d|\ti{\k}| \,     \ti{\eta}(\ti{\k}),  
\eeqa
 which is strictly positive. Since $b_n^{-2}\to \infty$, we obtain (\ref{to-verify-lightcone-norm}). Next, suppose $\r=0$. Then the integral in (\ref{strictly-positive-integral}) reads
 \beqa
 \int_{L/2}^{L}  d|\ti{\k}| \,  \ti{\eta}(\ti{\k}) \cos( |\ti{\k}| /L), \label{modified-discussion-end}
 \eeqa
 which is also strictly positive. Thus we conclude the proof of Theorem~\ref{no-lightcone-normality}.  

As we anticipated  in (\ref{W-F-R}),   the timelike order parameter diverges
to infinity in the disordered phase and at the critical point. Actually the same is true for the spacelike order parameter. {\red In fact,}
 the  discussion (\ref{modified-discussion-start})--(\ref{modified-discussion-end}) above can be immediately adapted to the
spacelike case by choosing $\tau=0$ and the support of $\eta$ outside  zero. Thus we have verified the second line in (\ref{distinction}).

{\red Interestingly}, {\crr for $\r>0$} the divergence of the spacelike order parameter can be seen directly from the existence of the $S$-matrix,
using the  method of central sequences from the proof of Lemma~\ref{central-lemma}.
Coming back to (\ref{S-matrix-computation}), we have
\beqa
\e^{-\i \mrm{Im}\lan \vv^{\mrm{out}}-  \vv^{\mrm{in}}, F_R\ran }\pi_{\r}(W(F_R)) \2=\2 S_{\r} \pi_{\r}(W(F_R))S_{\r}^*, 
\label{S-matrix-computation-one}
\eeqa
for a unitary $S_{\r}$. We choose here  $F_R$  as in (\ref{central-sequence-def}), but for $f$ supported in the spacelike complement of zero.
We compute
\beqa
\mrm{Im} \lan \vv^{\mrm{out}}-  \vv^{\mrm{in}}, F_{R} \ran 
\underset{R\to \infty}{\to}  \ti j_0(0) \int {\rtable d^3\k} \,   \fr{ \mrm{Re}(F(\k))}{ |\vk|^{3/2}} \bigg\{\fr{1}{(1-\hat{\k} \cdot \v^{\mrm{out}})} - \fr{1}{(1-\hat{\k} \cdot \v^{\mrm{in}})}  \bigg\},\label{coherent-x}
\eeqa
which can easily be arranged to be different from zero (and multiples of $2\pi$) for $\v^{\mrm{out}} \neq \v^{\mrm{in}}$. 
Now  arguing as in the proof of Lemma~\ref{central-lemma}, we obtain
\beqa
\lim_{R\to \infty}(\e^{-\i \mrm{Im}\lan \vv^{\mrm{out}}-  \vv^{\mrm{in}}, F_R\ran }-1)  \lan \Om, \pi_{\r}(W(F_R))\Om\ran=0,
\eeqa
which implies $\lan \Om, \pi_{\r}(W(F_R))\Om\ran\to 0$, hence  $\om_{\r}(\phi_R(f)\phi_R(f))\to \infty$, cf. (\ref{W-F-R}). Thus diverging 
fluctuations of the field at spacelike infinity are  an unavoidable price for stabilizing the Bloch-Nordsieck $S$-matrix, independent of
the detailed structure of the states $\om_{\r}$. 

Unfortunately, such a general argument fails for fluctuations of the timelike order parameter, as   (\ref{coherent-x}) is
inevitably zero in this case \cite{CD20}. {\red (Unlike the spacelike case, $F$ cannot be chosen real here)}. Thus for the divergence of the fluctuations of the field in future timelike directions we have to rely on
computations presented earlier in this section which depended on the particular structure of the states $\om_{\r}$. The same applies to the 
absence of lightcone normality.

\section{Lightcone normality in the ordered phase} \label{Lightcone-section}
\setcounter{equation}{0}

In this section we prove the following theorem:
\bet \label{lightcone-normality-thm} The representations $\pi_{\r}$ are lightcone normal for {\crr $\r<0$}. 
\eet
\nin Our proof is inspired by \cite{DFG84}, but we consider a more complicated family of states.  We use a general fact that any state $\om$ on $\mfa$ such that
\beqa
|\om(A)-\om_{\vac}(A)|<2\|A\|, \quad A\in \mfa(V_+),  \label{Haag-book-criterion}
\eeqa   
is lightcone normal, cf. \cite[Theorem 2.2.16]{Ha}.  To check this criterion, we will use representation~(\ref{def-om-r}) of  states $\om_{\r}$.  
We recall the operators $\mcUr(I_n)$ from  (\ref{Weyl-operator-implementation})  and generalize them to any subset $I\subset I_n$.
Now by the Weyl relations and the definition of the vacuum (\ref{Weyl+vacuum}) the approximating sequence of   $\om_{\r}$ from   (\ref{def-om-r}) can be restated
as follows: 
\beqa
& &\lan \mcUr(I)\Om, W(F)   \mcUr(I)\Om  \ran\non\\
& &=\bigg(\prod_{\al'} \fr{z_{\al'}}{2\pi} \bigg)  \int  d^{\nI}s \int d^{\nI}t\, \e^{- \hh \sum_{\al}(\ww_{\al}  -1) s_{\al}^2} 
 \e^{- \hh \sum_{\be} (\ww_{\be}  -1) t_{\be}^2}  \e^{-\fr{1}{4}\sum_{\al} (s_{\al}-t_{\al})^2} \label{representation-one} \\
& &\times \e^{\fr{\i}{2} [   t_{\al} \mrm{Im}\lan  \tfmal, F\ran+ s_{\be} \mrm{Im}\lan  \tbfmbe, F\ran    + {\mrm{Im}\lan F, - s_{\be} \tfmbe- t_{\be} \tbfmbe  \ran} ] }  \e^{\fr{1}{4} 2\mrm{Re} \lan F, \stw (s -t)_{\al} (\tbfmal-\tfmal) \ran   } \label{to-vanish-one}\\
&  &\times \e^{ \fr{1}{4}    [    s_{\al} s_{\be} \mrm{q}_{\al,\be} +t_{\al}t_{\be} \mrm{p}_{\al,\be}  - 2s_{\al} t_{\be}    \mrm{r}_{\al,\be}   ]   }  \label{q-p-r} \\ 
 & &\times \lan\Om,W(- t_{\al} d_{+,\al}+ s_{\al} d_{-,\al} )W(F) W( s_{\be} d_{+,\be} - t_{\be} d_{-,\be}  )\Om\ran, \label{representation-three}
\eeqa
where $F\in \mcL$, $f_{\al}\in D(\real^4;\complex)$, $\ti{f}_{\al,\mu}(\k):=\mu^{-1/2}(\k)\ti{f}_{\al}(|\k|,\k)$ and summation over repeated indices $\al, \be$ is understood.

Let us start the discussion of this lengthy formula from factor~(\ref{to-vanish-one}). Here, by (\ref{Haag-book-criterion}), it suffices to consider $F\in \mcL(V_+)$,
where $\mcL(V_+)=\{ \phi(f)\Om\,|\, f\in D(\real^4;\real),  \supp(f)\subset V_+\}$.  Restricting attention to $f_{\al}\in D(V_-;\complex)$, {\red where $V_-$ is the backward lightcone,} this factor is equal to one by the Huyghens principle. This is a significant step towards verifying (\ref{Haag-book-criterion}), as now  function 
$F$ appears only in  (\ref{representation-three}) and both sides of the equation extend to linear functionals acting on $A\in \mfa(V_+)$.

Regarding factors (\ref{q-p-r}), (\ref{representation-three}), the rule of the game is to choose $f_{\al}$ in such a way that the quantities 
\beqa
d_{+,\al}:= \i\hzeta_{\al}-\tfmal, \ \ d_{-,\al}:=\tbfmal \label{d-definition}
\eeqa
are small.  This will allow us, in particular, to control the factor (\ref{q-p-r}), since the quantities
\beqa
\mrm{q}_{\al,\be}:= w_{\al,\be}-2\i  u^{+,-}_{\al,\be}, \quad
\mrm{p}_{\al,\be}:=w_{\al,\be} +2\i  u^{+,-}_{\al,\be}, \quad
\mrm{r}_{\al,\be}:=  \mrm{Re}(w_{\al,\be}){\crr -} \i( u^{+,+}_{\al,\be} - u^{-,-}_{\al,\be})
\eeqa
depend on $d_{\pm}$ via
\beqa
 \ \ w_{\al,\be}:=\lan d_{+,\al}+d_{-,\al},   d_{+,\be}+d_{-,\be} \ran, \ \
u^{\si_1,\si_2}_{\al,\be}:= \mrm{Im}\lan d_{\si_1,\al}, d_{\si_2,\be} \ran, \quad \si_1,\si_2\in \{\pm\}.
\eeqa
Now if we find such $f_{\al}$ that $d_{\pm,\al}\to 0$ then the r.h.s. of  identity (\ref{representation-one})--(\ref{representation-three}) approaches the vacuum state as needed to check criterion (\ref{Haag-book-criterion}). 

Let us indicate how to control the resulting
error terms:  First, we note the straightforward estimate
  \beqa
   & &|t_{\al}  \,\mrm{q}_{\al,\be} s_{\be}|, |t_{\al}  \,\mrm{p}_{\al,\be} s_{\be}|,  |t_{\al}  \,\mrm{r}_{\al,\be} s_{\be}|\leq 
   \fr{3}{2} \vertiii{d}^2  \sum_{\al} (|t_{\al}|^2+|s_{\al}|^2  )\|d_{\al}\|_2,
  \eeqa
 where  we set $\|d_{\al}\|_2:=\|d_{+,\al}\|_2+\|d_{-,\al}\|_2$, $\vertiii{d}:=\sum_{\al} \|d_{\al}\|_2$ and applied the Cauchy-Schwarz inequality.
 Thus we obtain the following bound for error terms originating from (\ref{q-p-r}): 
 \beqa
|  \e^{ \fr{1}{4}    [    s_{\al} s_{\be} \mrm{q}_{\al,\be} +t_{\al}t_{\be} \mrm{p}_{\al,\be}  - 2s_{\al} t_{\be}    \mrm{r}_{\al,\be}   ]   }  -1|\leq 
\vertiii{d} \e^{3\sum_{\al} (|t_{\al}|^2+|s_{\al}|^2  )\|d_{\al}\|_2 }, \label{universal-bound}
\eeqa
where we noted that $|\e^{x}-1| \leq y  \e^{(1+y^{-1}) |x|}$ for $x\in \real, y>0$, 
and assumed that $ \vertiii{d} \leq 1$. By functional calculus, the same bound can be established for quantities
 \beqa
 \|\big(W( - t_{\al} d_{+,\al}+ s_{\al} d_{-,\al} )^* -1\big)\Om\|, \quad \|\big(W( s_{\be} d_{+,\be} - t_{\be} d_{-,\be} ) -1\big)\Om\|,
\eeqa
 which appear in error terms originating from (\ref{representation-three}). Altogether, using (\ref{universal-bound}) and (\ref{representation-one})--(\ref{representation-three})  we easily obtain the following bound 
 \beqa
 \2 \2|\lan \mcUr(I)\Om, A   \mcUr(I)\Om  \ran- \lan \Om, A\Om\ran| \non\\ 
 \2 \2\leq\vertiii{d} \|A\|\bigg(\prod_{\al} \fr{z_{\al}}{2\pi} \bigg)  \int  d^{\nI}s \int d^{\nI}t\, 
\e^{- \hh \sum_{\al}([\ww_{\al}-12\|d_{\al}\|_2]  -1) s_{\al}^2} 
  \e^{- \hh \sum_{\be} ([\ww_{\be}-12\|d_{\be}\|_2] -1) t_{\be}^2}  \e^{-\fr{1}{4}\sum_{\al} (s_{\al}-t_{\al})^2} \non\\
\2 \2= \vertiii{d}\|A\| \bigg(\prod_{\al} \fr{z_{\al}}{\ti{z}_{\al}} \bigg), \label{z-quotient-zero}
 \eeqa
where   $z_{\al}:= \h\sqrt{\ww_{\al}^2-1} $  as defined before, $\ti{z}_{\al}:= \h\sqrt{[\ww_{\al}-12\|d_{\al}\|_2 ]^2-1}$ and in the last step we used
Gaussian integration.  Now by elementary estimates we arrive at the following lemma:
 
\bel\label{closeness-to-vacuum}  Let $\vertiii{d}\leq 1$.  Suppose that $\ww_{\al}-12\|d_{\al}\|_2-1>0$ for $\al\in I$. 
Then the following estimate holds
\beqa
|\lan \mcUr(I)\Om, A   \mcUr(I)\Om  \ran- \lan \Om, A\Om\ran|\leq c\|A\| \vertiii{d} 
 \exp\bigg(c'\sup_{\be\in I}\ww_{\be}\sum_{\al\in I} \fr{ \|d_{\al} \|_2}{\ww_{\al}-12 \|d_{\al}\|_2 -1}\bigg) \label{estimate-for-Haag-bound}
 \quad 
\eeqa
for $A\in \mfa(V_+)$ and some numerical constants $c$, $c'$.
\eel
 The next task is to find   $f_{\al}\in D(V_-; \complex)$ such that $d_{\pm,\al}\to 0$ sufficiently fast,
given our set of timelike shifts $\{\taur_{\al}\}_{\al\in I_n}$ of (\ref{taur-i}). With such input, we intend to estimate
the r.h.s. of  (\ref{estimate-for-Haag-bound}) so as to obtain (\ref{Haag-book-criterion}).
 As a first step, in  Appendix~\ref{approximation-problem}
we find for any  $\ti r>0$ a family of functions $f^{\circ}_i\in D(\real^4;\complex)$ supported in double cones of radii {\crr $2/\de_i$,}
$\de_i:={\crr 2} \eps_{i}^{1+\ti r}$,  such that
\beqa
\| \xii_i {\crr \i Y_{\ell m}}- \ti{f}_{\mu,i}^{{\crr\circ}}\|_2\leq c_{\ti r} i^2\eps_i^{\ti r/4},\quad  \| \ti{\bar{f}}_{\mu}^{{\crr\circ}}  \|_2 \leq c_{\ti r} i^2\eps_i^{\ti r/4}, 
\label{two-differences-zero}
\eeqa
where the constants $c_{\ti r}$ are independent of $i$.  
Next, we assume that $\ti r\leq |r|$ and  define the functions appearing in (\ref{d-definition})  as $\ti{f}_{\mu,i}:= \e^{-\i \mu {\crr \taur_i}  } \ti{f}_{\mu,i}^{{\crr\circ}}$.
Clearly,  the fields  $\phi(f_i)$   are localized in the backward lightcone ${\red V_-}$. As a matter of fact, the bounds (\ref{two-differences-zero}) are at the basis of the introductory discussion in {\red Sub}section~\ref{description-transition} and the double cones in Fig.~1  symbolize the localization regions of the fields $\phi(f_i)$ in the case $0<\ti r \ll |r|$.   By invariance of the norm under the unitary time evolution, the  bounds (\ref{two-differences-zero}) hold also for the quantities
\beqa
\|d_{+,\al}\|_2=\| \e^{-\i \mu {\crr \tau_i}  } \xii_i {\crr \i Y_{\ell m}} -  \e^{-\i \mu {\crr \tau_i}  } \ti{f}_{\mu,i}^{{\crr\circ}}\|_2
,\quad  
\|d_{-,\al}\|_2=\| \e^{-\i \mu {\crr \tau_i}  }  \ti{\bar{f}}_{\mu}^{{\crr\circ}}  \|_2. \label{d-quantities}
\eeqa 
 
 After this preparation we are almost ready to prove Theorem~\ref{lightcone-normality-thm}. The last remaining difficulty is to obtain from (\ref{estimate-for-Haag-bound})
 the bound by $2\|A\|$ appearing in (\ref{Haag-book-criterion}). The problem is that $d_{\pm,\al}\to 0$  does not mean that  $\vertiii{d}:=\sum_{\al} \|d_{\al}\|_2$,
 appearing on the r.h.s. of (\ref{estimate-for-Haag-bound}), is small. In fact, although the sequence {\red $\|d_{\al}\|_{2}$} converges to zero, first few terms can be large. However, the smallness of $\vertiii{d}$ can easily be ensured by dropping a finite number
 of modes from the states $\om_r$. (This amounts to dropping a finite number of terms in the sums in (\ref{T-r-def})). It is intuitively clear  that this leads to a unitarily equivalent representation. After all, {\red by the Stone-von Neumann uniqueness theorem,} the appearance of distinct sectors requires infinitely
 many degrees of freedom, cf.  Appendix~\ref{omitting} for details.  Thus we conclude the proof of Theorem~\ref{lightcone-normality-thm}  using  Lemma~\ref{closeness-to-vacuum} and estimate (\ref{Haag-book-criterion}) given bounds (\ref{two-differences-zero}) which hold also for (\ref{d-quantities}). 

\section{Order parameters in the ordered phase} \label{ordered-section}
\setcounter{equation}{0}

In this section we determine the timelike and spacelike order parameters in the ordered phase. {\red Thus  we} verify 
the first line in (\ref{distinction}), which we left aside so far. Regarding the  timelike case, we can use the lightcone normality of $\pi_{\r}$:
{Since the representation is equivalent to the vacuum, it is not surprising that the timelike asymptotic fluctuations of the field take the vacuum value}.
{\red At the technical level,} this statement follows immediately from Lemma~\ref{central-lemma}. The spacelike case is more interesting as
we cannot conclude the vacuum form of the order parameter directly from lightcone normality.
However, some ingredients from the proof of lightcone normality {\red will also prove} important here.

Since we are looking at the spacelike order parameter, we can choose such $f$ in (\ref{central-sequence-def})  that the resulting  $F$  is  real-valued. 
Then we obtain from formula~(\ref{T-r-def})
\beqa
\om_{\r}(\phi_R(f)  \phi_R(f))\2 =\2\|\Tr F_{R}\|^2_2\non\\
\2=\2\|F\|_2^2+\sum_{\al\in I_{\infty}} \big(b_{\al}^{-2} -1\big)    |  \mrm{Re} \lan \hzeta_{\al} ,F_{R}\ran|^2 
+\sum_{\al\in I_{\infty}} \big(b_{\al}^2 -1\big)  |\mrm{Im}\lan \hzeta_{\al},  F_{R}\ran|^2. \label{last-sum}
\eeqa
Using functions $f_{\al}$, which appeared in (\ref{d-definition}), we 
rewrite the above scalar products as follows:
\beqa
\lan \hzeta_{\al}, F_{R}\ran=\lan \hzeta_{\al}-\ti{f}_{\al,\mu}, F_{R}\ran+\lan \ti{f}_{\al,\mu}, F_{R}\ran. \label{two-terms-last-section}
\eeqa 
Now recalling (\ref{two-differences-zero}), (\ref{d-quantities}), the first term satisfies 
\beqa
|\lan \hzeta_{\al}-\ti{f}_{\al,\mu}, F_{R}\ran| \leq \|F\|_2c_{\r} i^2 \eps_i^{ {\crr |\r|} /4},  \label{shift-estimate}
\eeqa
where we chose $\ti{r}=|r|$.
As for the second term in (\ref{two-terms-last-section}), let us first restrict attention to  $\al$, $R$ such that $\phi(f_\al)$ and $\phi_{R}(f)$ commute.
Then
\beqa
|\lan \ti{f}_{\al,\mu}, F_{R}\ran|= |\lan \phi(f_{\al})\Om, \phi_{R}(f)  \ran | {\red =}  |\lan \phi_{R}(f) \Om, \phi(\bar{f}_{\al})\Om\ran |\leq \|F\|_2\| \ti{\bar{f}}_{\al, {\red \mu}}\|_2
\leq \|F\|_2c_{\r} i^2 \eps_i^{{\crr |\r|} }, \label{locality-estimate}
\eeqa
where we used again (\ref{two-differences-zero}), (\ref{d-quantities}). 

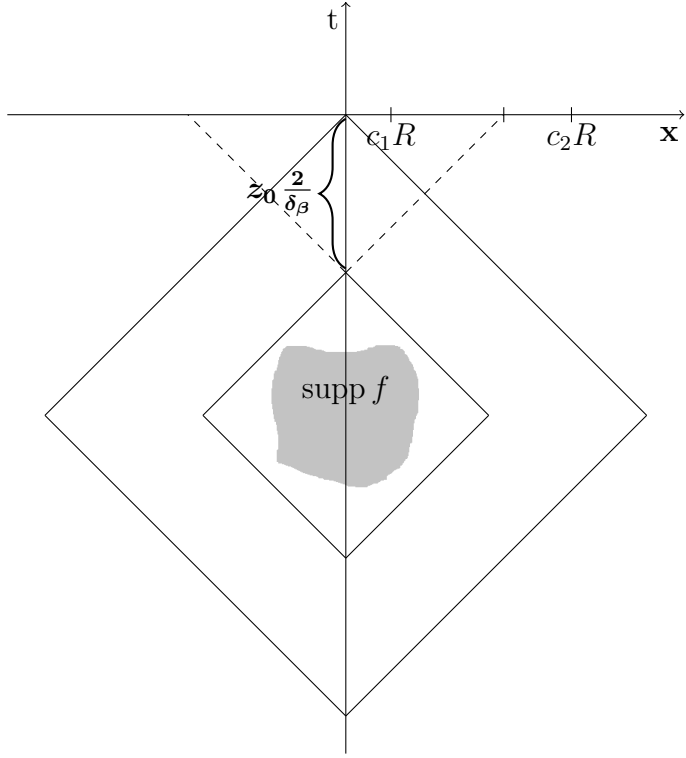
\begin{figure}[t]
\centering
\begin{tikzpicture}[scale=1][h]
	\node[inner sep=0pt][scale=0.4] at (0,0) {\includegraphics[width=5cm]{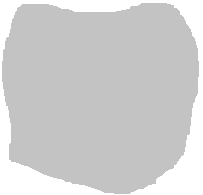}};
	\draw (0,0) node[anchor=south] {$\operatorname{supp}f$};
	\draw (0,-4) -- (4,0);
	\draw[] (4,0) -- (0,4);
	\draw (0,-4) -- (-4,0);
	\draw[] (-4,0) -- (0,4);
	\draw (0,-1.9) -- (+1.9,0);
	\draw (+1.9,0) -- (0,+1.9);
	\draw (0,-1.9) -- (-1.9,0);
	\draw (-1.9,0) -- (0,+1.9);
	\draw[dashed] (0,+1.9) -- (0-2.1,+1.9+2.1);
	\draw[dashed] (0,+1.9) -- (0+2.1,+1.9+2.1);
	\draw[->] (-4.5,4) -- (4.5,4);
	\draw[->] (0,-4.5) -- (0,5.5);
	\draw (4.3,4-0.5) node[anchor=south] {\textbf{x}};
	\draw (-0.17,+5) node[anchor=south] {t};
	\draw (2.1,4-0.1) -- (2.1,4+0.1);
	\node (A) at (0,3.8) {};
	\node (B) at (0,2.1) {};
	\draw[decorate,decoration={brace,amplitude=10pt,mirror},thick] (A.north) -- (B.south) node[midway,xshift=-25pt] {{\rtable $\mathbf{z_{0} \hspace{0.05cm} \frac{2}{\delta_{\beta}}}$}};
	\draw (0.6,4-0.1) -- (0.6,4+0.1);
	\draw (3.0,4-0.1) -- (3.0,4+0.1);  
	\draw (3.0,4-0.61) node[anchor=south] {$c_{2}R$};   
	\draw (0.6,4-0.61) node[anchor=south] {$c_{1}R$};
\end{tikzpicture}
\caption*{\footnotesize{Fig.~3. Schematic illustration for relation~(\ref{support-relations}).} }
\end{figure}

Now we determine the set of $\al$, $R$ for which the commutation of   $\phi(f_\al)$ and $\phi_{R}(f)$ may fail. We recall from the discussion above (\ref{two-differences-zero}),  that the function $f^{\circ}_{\al}$,
for $\al=(1,0,0)$, was supported in the double cone of radius $1$ centered at zero. Since the support
 of a function is compact and the double cone is an open set, {\red this function is} also supported in a double cone of a slightly smaller radius $(1-z_0)$. 
 For arbitrary $\al$, functions $ f^{\circ}_{\al}$   are supported in  double cones of radii $\fr{2}{\de_{\al}}(1-z_0)$, where $z_0$ is independent of 
 $\al$. This is due to the fact that these supports  simply arise  by scaling the supports of certain auxiliary functions $\ti{\eta}$, $\ti{\eta}_1$, cf. Appendix~\ref{approximation-problem}.    
Taking the  timelike shifts into account, the support of  $f_\al$ {\red is contained in}
\beqa
\mco_{ \fr{2}{\de_{\al}}(1-z_0)} - e_0\taur_{\al}=\mco_{ \fr{2}{\de_{\al}}(1-z_0)} -e_0\fr{2}{\de_{\al}},
\eeqa
where $e_0$ is the timelike unit vector {\red and $\mco_a$ is the double cone of radius $a$ centered at zero}.  Now suppose that $f$ (entering into the order parameter) is supported in the annulus 
$0<c_1< |x|< c_2$. Then $\phi_{R}(f)$ and $\phi(f_{{\red \al}})$ commute, unless 
\beqa
z_0\fr{2}{\de_{\al}} < c_2 R 
\quad\Rightarrow\quad \fr{z_0}{ 2c_2} (2^{{\crr |\r|}})^{i-1}<  \eps_{i+1}R, \label{support-relations}
\eeqa
where we recalled that $\de_{\al}=2\eps_{\al}^{{\crr 1+|\r|}}$.  This can be read off from Fig.~3, where  $\ti r=|r|$ was chosen  (unlike Fig.~1 drawn for $\ti r\ll |r|$). 
For $\al$ as in (\ref{support-relations}) we cannot argue as in  (\ref{locality-estimate})
but instead we get a similar bound on the quantity in (\ref{two-terms-last-section}) by an explicit computation: Choosing $f$ depending only on $\x$ and spherically symmetric, we obtain 
\beqa
|\lan \hzeta_{\al}, F_{R}\ran| 
\2=\2 (4\pi) \big| \int^{\eps_i R  }_{\eps_{i+1}R} d|\k'| \, \e^{\i   |\k'| \taur_i  R^{-1} }  \ti{f}(|\k'|)\big| \non\\
\2\leq\2 (4\pi) \int_{\fr{z_0}{ 4c_2} (2^{{\crr|\r|}})^{i-1}}^{\infty} d|\k'| \,| \ti{f}(|\k'|)| \leq c_N 
[(2^{ {\crr |\r|} })^{i-1}]^{-N}, \label{no-restriction}
\eeqa
where in the last two steps we made use of (\ref{support-relations}) and of {\red the} rapid decay of $f$.  Now the discussion  
(\ref{two-terms-last-section})--(\ref{no-restriction})  gives 
\beqa
|\lan \hzeta_{\al}, F_{R}\ran|\leq c'_{\r} i^2 \eps_i^{{\crr |\r|} }, \label{summarizing-estimate}
\eeqa
for some constant $c'_{\r}$ independent of $R$, ${\red \al}$. {\red There are no restrictions on $R$, $\al$ anymore, as we treated all the possibilities.}

Finally, using (\ref{summarizing-estimate}),  we obtain  bounds on the summands in (\ref{last-sum}) which are independent of $R$ and summable.
Thus, by the dominated convergence theorem, we can enter with the limit $R\to \infty$ under the sums. Then it is clear from the first step in (\ref{no-restriction}) that the
individual summands tend to zero and we are left with the vacuum contribution $\|F\|_2^2$.

\section{Conclusion and outlook} \label{Outlook}

In this paper we described a family of states $\{\om_{\r}\}_{\r\in \real}$ interpolating between the vacuum  at {\rtable $\r\to -\infty$} and
the Kraus-Polley-Reents infravacuum  at {\crr $\r\to \infty$}. We found a phase transition at $\r=0$ separating  the
ordered phase {\crr ($\r<0$)} from the disordered phase {\crr ($\r>0$)}. Remarkably, in the ordered phase the state is lightcone normal, that is,
it cannot be distinguished by any experiments in the future lightcone from a state in the vacuum sector. Not surprisingly, this
phase suffers from all the infrared problems  familiar from the vacuum sector: the Bloch-Nordsieck $S$-matrix disintegrates and
soft photon clouds break rotational symmetry. However, by an infinitesimal change of the control parameter from {\crr $\r={\red -\epsilon}$ to $\r={\red+\epsilon}$} 
we enter the disordered phase, where the $S$-matrix is stabilized by the Kraus-Polley-Reents mechanism and rotational symmetry {\rtable is} restored.
The transition can be observed in  the future lightcone by measuring the timelike asymptotic fluctuations of the field. {\red Their abrupt increase prevents the
lightcone normality in the disordered phase.}

Our phase transition is quite subtle and {\red does not easily  fit into standard categories.}  As states $\om_{\r}$ are not ground states, 
{\red it is} a non-equilibrium phase transition, which consists in a qualitative change of a certain dynamics. This dynamics  is,
in our case, the Bloch-Nordsieck scattering process and the qualitative change consists in {\red the existence or disintegration of its scattering matrix}.  
 This transition is {\red also} visible in the asymptotic  fluctuations of the order parameter
stated in (\ref{distinction}), which  have a discontinuity at $\r=0$. On the one hand, such discontinuity suggests a first order phase transition. {\rtable On the other hand,}
the behaviour of the correlation length $\lan \Om, S_{\r}\Om\ran^{-1}\sim \exp(c / \r^{3/2})$,  stated in (\ref{distinction-S}), points to a 
continuous transition.  It would be interesting to study the disintegration of other quantities near $\r=0$, such as  the  angular momentum  in 
representations $\pi_{\r}\circ \alpha$. It would also be desirable to identify some universal features of the resulting dependencies (`critical exponents')
independent of the detailed  construction of the states $\om_{\r}$.  Finally, the transition is clearly visible in {\red the boundary} degrees of freedom, 
 but it may be accompanied by some measurable effects in the bulk. For example, the rate of decay of correlations 
$\x\mapsto \om_{\r}(\phi(0)\phi(\x))$ might distinguish the two phases. We leave these questions to future research.

The Kraus-Polley-Reents mechanism for curing infrared problems has obvious advantages over the usual infraparticle (or Faddeev-Kulish)
approach. Instead of intricate, velocity dependent dressing transformations of the $S$-matrix,  it provides a  regularizing background radiation which is
independent of  the dynamics. After replacing the vacuum with a KPR infravacuum the {\red usual Dyson $S$-matrix is meaningful}. This is 
clear  in the external source situation, but also thinkable in perturbation theory. Apart from the scalar field, the KPR infravacua are available for the electromagnetic field \cite{KPR77, CD18} and we are confident that our findings could be generalized to this case. However, to our knowledge the KPR mechanism has not yet been tested
in the case of  massless higher spin particles. It would be interesting to fill this gap,  given the difficulties with applying the Faddeev-Kulish approach
in the presence of gravitons \cite{PSW22}. In the next step one could revisit the approach to the black hole information paradox  from \cite{HPS16} in the
presence of  KPR radiation. As such radiation dramatically changes the superselection structure of infrared degrees of freedom, {\red this} may lead to interesting new insights.

\appendix

\section{Outline of the proof of  estimates (\ref{two-differences-zero})} \label{approximation-problem}
\setcounter{equation}{0}

\newcommand{\elli}{(\ell+1)}

For brevity, we skip the index ${\crr \al}$ on $f$, $\xii$, $\taur$  and write $Y:=Y_{\ell m}$. We consider $f^{\circ}\in D(\real^4;\complex)$ of the form $f^{\circ}(t,{\rtable \x})=f_0(t)f_1({\rtable \x})$. We require that it is supported in a double cone of radius $\fr{2}{\de}$, for some  $0<\de\leq 2$. We have 
\beqa
\ti{f}^{{\crr\circ}}_{\mu}(\k)\2=\2   |\k|^{-1/2} \fr{1}{(2\pi)^{\crr 3/2} } \int dt {\rtable d^3\x} \, \e^{\i |\k|t} \e^{-\i \k\cdot \x} f_0(t) f_1(\x) = |\k|^{-1/2} \tilde{f}_0(|\k|) \tilde{f}_1(\k), \label{f-mu} \\
\ti{\bar{f}}^{{\crr\circ}}_{\mu}(\k)\2=\2 |\k|^{-1/2}\bar{\tilde{f}}_0(-|\k|) \bar{\tilde{f}}_1(-\k). 
\eeqa  
 We will choose $f_0(t)=\bar{f}_0(-t)$ so that $\tilde{f}_0$ is real.
 We want to choose $\tilde{f}_0(k^0)\simeq 1$ for $k^0>0$ and $\tilde{f}_0(k^0)\simeq 0$ for $k^0<0$. For this purpose, we set
\beqa
\tilde{f}_0(k^0):=(\eta_{0,\de}*\chi_{[0,2]})(k^0), \label{construction-function-appendix}
\eeqa
where $\eta_{0,\de}(\,\cdot\,):=\fr{1}{\de} \eta_{\crr 0}(\fr{1}{\de}(\,\cdot\,))$ and $\eta_{\crr 0}\in D(\real;\real)$ is positive, symmetric and such that   $\int dk^0\, \eta_0(k^0)=1$ so that $\eta_{0,\de}$ is a delta approximating sequence as $\de\to 0$.   We also require that the Fourier transform of $\eta_{\crr 0}$ is supported in 
{\crr the interior of} a ball of radius one so that $\ti{\eta}_{\de}$  is supported inside a ball of radius $1/ \de$.   As for $f_1$,
we choose analogously a delta approximating sequence $\eta_{1,\de}(\,\cdot\,):=\fr{1}{\de^3} \eta_{1}(\fr{1}{\de}(\,\cdot\,))$ in momentum space, such that its  Fourier transform is compactly supported in the interior of a ball of radius $1/\de$. We  set
\beqa
\ti{f}_1(\k):=  (\eta_{1,\de}* \mu^{1/2} {\xii}{\crr \i Y}  )(\k).
\eeqa

By slightly tedious, but straightforward estimates one checks that for any $\ti r>0$ and $\de:={\crr 2} \eps_{i}^{1+\ti r}$  
\beqa
\| \xii {\crr \i Y}- \ti{f}_{\mu}^{{\crr\circ}}\|_2\leq c_{\ti r} i^2\eps_i^{\ti r/4},\quad  \| \ti{\bar{f}}_{\mu}^{{\crr\circ}}  \|_2 \leq c_{\ti r} i^2\eps_i^{\ti r/4}, \label{two-differences}
\eeqa
where the constants $c_{\ti r}$ are independent of $i$. The essential ingredients here are the rapid decay of the functions $\eta_0,\eta_1$,
the uniform bound  $|\ti{f}_0(k^0)| \leq 1$ and basic properties of spherical harmonics:
\beqa
|Y_{\ell m}(\hat{\k})| \2\leq\2 c(\ell+1)^{1/2},  \label{s-h-bound-zero} \\  
|Y_{\ell m}(\hat{\k}) -Y_{\ell,m}(\hat{\k}')| \2\leq\2 \ti{\alpha} \ell(\ell+1), \label{spherical-shift-two-zero}
\eeqa
where $\ti \alpha$ is the angle between the unit vectors $\hat{\k},  \hat{\k}'$.
For (\ref{s-h-bound-zero})  we refer to \cite[formula (2.36)]{AH}. Estimate (\ref{spherical-shift-two-zero}) 
follows by expressing the rotation from $\hat{\k}$ to $\hat{\k'}$ in terms of the angular momentum operators.
The factor $i^2$ on the r.h.s. of (\ref{two-differences}) comes from estimates (\ref{s-h-bound-zero}), (\ref{spherical-shift-two-zero})
and the condition $\ell<i$ appearing in (\ref{pmbQi}). 
\section{Omitting finite number of modes from $\om_r$} \label{omitting}
\setcounter{equation}{0}

We provide here some details for the last step of the proof of Theorem~\ref{lightcone-normality-thm}.
We define $I^{n_0}_n:=\{\, \al=(i,\ell,m) \in I_n\,|\, i>n_0\}$ for $n_0<n$
and decompose $I_{n}=I_{n_0}\cup I^{n_0}_n$.  By  the arguments which led to (\ref{om-r-def-one}), (\ref{om-r-def-two}) we can write
\beqa
\lan \mcUr(I^{n_0}_n)\Om, W(F) \mcUr(I^{n_0}_n)\Om{\red \ran}=\lan \Om, W(\T_{r,I^{n_0}_n } F) \Om\ran,
\eeqa
where the symplectic map $\T_{r,I^{n_0}_n}$ arises by keeping only $n_0<i\leq n$ in (\ref{T-r-def}).  Denoting by $\pi_{\r,I}$ the representation 
acting by $\pi_{\r,I}(W(F))=W(\T_{r,I}F)$, we note the relation
\beqa
\lan\Om, \pi_{\r, I_n} (W(F))\Om\ran=\lan \mcUr(I_{n})\Om, W(F) \mcUr(I_{n})\Om\ran=\lan \mcUr(I_{n_0})\Om, \pi_{\r, I^{n_0}_n}(W(F)) \mcUr(I_{n_0})\Om\ran.
\eeqa
By keeping $n_0$ fixed and taking the limit $n\to \infty$, we obtain
\beqa
\lan\Om, \pi_{\r} (W(F))\Om\ran= \lan \mcUr(I_{n_0})\Om, \pi_{\r,  I^{n_0}_{\infty} }(W(F)) \mcUr(I_{n_0})\Om\ran.
\eeqa
Thus, by \cite[Lemma A.1]{CD18}, $\pi_{\r}$ is unitarily equivalent to $\pi_{\r,  I^{n_0}_{\infty} }$ for any $n_0\in \nat$.



\begin{thebibliography}{LNT2}


\bibitem[Ar]{Ar} A. Arai. \emph{Inequivalent representations of canonical commutation and anti-commutation relations.} Springer, 2020.  

\bibitem[AY82]{AY82} H. Araki and S. Yamagami. \emph{On quasi-equivalence of quasifree states of the canonical
commutation relations}. Publ. RIMS, Kyoto Univ. \textbf{18}, (1982) 283--338.  

\bibitem[AS81]{AS81} A. Ashtekar and M. Streubel. \emph{Symplectic geometry of radiative modes and conserved quantities at null infinity}. Proc. R. Soc. Lond. A 376, (1981) 585–607.

\bibitem[AH]{AH} K. Atkinson and W. Han. \emph{Spherical Harmonics and Approximations
on the Unit Sphere: an Introduction.} Lecture Notes in Mathematics, Springer, 2012.


\bibitem[BD22]{BD22} B. Biadasiewicz and   W. Dybalski.  \emph{Local normality of infravacua and relative normalizers for relativistic systems}. Lett.  Math. Phys.  \textbf{112},  (2022) 40. 





\bibitem[BG13]{BG13} L. Bieri and D. Garfinke. \emph{An electromagnetic analog of gravitational wave memory}.
Class. Quantum Grav. \textbf{30}, (2013) 195009.

\bibitem[BN37]{BN37} F. Bloch and A. Nordsieck. \emph{Note on the Radiation Field of the Electron}. Phys. Rev. \textbf{52}, (1937) 54--59.



\bibitem[Br77]{Br77} B.D. Bramson. \emph{Physics in cone space.} In: F.P. Esposito, E. Witten. (eds.) Asymptotic Structure of Space-Time, pp. 273–359. Plenum Press, New York (1977).


\bibitem[BR]{BR} O. Bratteli and D.W. Robinson. \emph{Operator algebras and quantum statistical mechanics I}. Springer, 1987.



\bibitem[Bu86]{Bu86} D. Buchholz. \emph{Gauss' law and the infraparticle problem}.
Phys. Lett. B  \bf 174\rm, (1986) 331--334.



\bibitem[BR14]{BR14} D. Buchholz and J.E. Roberts. \emph{New light on infrared problems: sectors, statistics, symmetries and spectrum.} Commun. Math. Phys. \bf 330\rm, (2014) 935--972.


\bibitem[CD19]{CD18} D. Cadamuro and W. Dybalski. \emph{Relative normalizers of automorphism groups, infravacua and the problem of velocity superselection in QED}.  Commun. Math. Phys. \textbf{372},  (2019) 769--796. 

\bibitem[CD20]{CD20}  D. Cadamuro and W. Dybalski. \emph{Curing velocity superselection in non-relativistic QED by restriction to a lightcone}. Ann. Henri  Poincar\'e  \textbf{21}, (2020) 2877--2896. 

\bibitem[CL15]{CL15} M. Campiglia and A. Laddha.  \emph{Asymptotic symmetries of QED and Weinberg's soft photon theorem}. JHEP\textbf{07} (2015) 115.

\bibitem[CE17]{CE17} M. Campiglia and R. Eyheralde. \emph{Asymptotic U(1) charges at spatial infinity}. JHEP\textbf{11} (2017) 168. 

\bibitem[CLPW22]{CLPW22} V. Chandrasekaran, R. Longo, G. Penington and E. Witten. 
\emph{An Algebra of Observables for de Sitter Space}. JHEP\textbf{02} (2023) 082.

\bibitem[Di97]{Di97} H.W. Diehl. \emph{The theory of boundary critical phenomena}.
International Journal of Modern Physics B \textbf{11} (1997) 3503--3523.


 
\bibitem[DFG84]{DFG84} S. Doplicher, F. Figliolini and D. Guido. \emph{Infrared representations of free Bose fields}. Ann. Inst. Henri Poincar\'e, \textrm{41}, (1984) 49--62.


\bibitem[DH19]{DH19} W. Dybalski and D. V. Hoang. \emph{A soft-photon theorem for the Maxwell-Lorentz system}. 
J. Math. Phys. \textbf{60}, (2019) 102903. 

\bibitem[DW19]{DW19} W. Dybalski and B. Wegener. \emph{Asymptotic charges, large gauge transformations and inequivalence of different gauges in external current QED}. JHEP\textbf{11} (2019) 126. 

\bibitem[GS16]{GS16} B. Gabai and A. Sever. \emph{Large gauge symmetries and asymptotic states in QED}. JHEP12 (2016) 095.

\bibitem[Ha]{Ha} R. Haag. \emph{Local quantum physics.} Springer,1996. 

\bibitem[HK64]{HK64} R. Haag and D. Kastler. \emph{An Algebraic Approach to Quantum Field Theory}. J. Math. Phys. \textbf{5},  (1964)  848--861. 

\bibitem[HPS16]{HPS16} S. W. Hawking, M. J. Perry and A. Strominger. \emph{Soft hair on black holes}.  Phys. Rev. Lett. \textbf{116}, (2016) 231301. 

\bibitem[HMPS14]{HMPS14}   T. He, P. Mitra, A.P. Porfyriadis and A. Strominger.   
\emph{New symmetries of massless QED}. JHEP\textbf{10} (2014) 112.

\bibitem[He16]{He16} A. Herdegen. \emph{Asymptotic structure of electrodynamics revisited}. Lett. Math. Phys.  \textbf{107}, (2017) 1439--1470.

\bibitem[He95]{He95} A. Herdegen. \emph{Long-range effects in asymptotic fields and angular momentum of classical field electrodynamics}. J. Math. Phys. \textbf{36}, (1995) 4044–4086.

\bibitem[Hi06]{Hi06} H. Hinrichsen. \emph{Non-equilibrium phase transitions}.
Physica A \textbf{369}, (2006) 1--28.




\bibitem[HSSS12]{HSSS12} F. Hiroshima, I. Sasaki, H. Spohn and A. Suzuki. \emph{Enhanced binding in quantum field theory.} 
Kyushu University COE Lecture Note \textbf{38}, 2012.


\bibitem[HIW16]{HIW16} S. Hollands, A. Ishibashi and R.M. Wald. \emph{BMS supertranslations and memory in four and higher dimensions}. Classical and Quantum Gravity \textbf{34}, (2017) 155005.



\bibitem[KR]{KR} R. V. Kadison and J. R. Ringrose. \emph{Fundamentals of the theory of operator algebras: Advanced theory.} Academic Press, 1986.

\bibitem[KSMLSC22]{KSMLSC22} M. Kalinowski, R. Samajdar, R. Melko, M.D. Lukin, 
S. Sachdev and S. Choi. \emph{Bulk and boundary quantum phase transitions in a square Rydberg atom array}. Phys. Rev. B \textbf{105}, (2022) 174417.


\bibitem[KPRS17]{KPRS17} D. Kapec, M. Perry, A.-M. Raclariu and A. Strominger.  \emph{Infrared divergencies in QED, revisited}. Phys. Rev. D \textbf{96}, (2017) 085002.



\bibitem[KPR77]{KPR77} K. Kraus, L. Polley and G. Reents. \emph{Models for infrared dynamics. I. Classical currents.} Ann. Inst. H. Poincar\'e  \textbf{26}, (1977) 109--162.

\bibitem[Kr82]{Kr82} K. Kraus.
 \emph{Aspects of the infrared problem in quantum electrodynamics.} Found. Phys. \textbf{13}, (1983) 701--713.

\bibitem[Ku98]{Ku98}  W. Kunhardt. \emph{On infravacua and the localisation of sectors}.  J. Math. Phys. \textbf{39},  (1998) 6353.


\bibitem[MRS22]{MRS22} J. Mund, K.-H. Rehren and B. Schroer. \emph{Infraparticle quantum fields and the formation of photon clouds}. JHEP\textbf{04} (2022) 083.

\bibitem[Pa17]{Pa17} S. Pasterski. \emph{Asymptotic symmetries and electromagnetic memory}. JHEP\textbf{09}  (2017) 154.

\bibitem[PSW22]{PSW22} K. Prabhu, G. Satishchandran and R.M. Wald. \emph{Infrared finite scattering theory in quantum field theory and quantum gravity}.
Phys. Rev. D \textbf{106}, (2022)  066005.

\bibitem[RS2]{RS2} M. Reed and B. Simon.\emph{Methods of Modern Mathematical Physics II. Fourier Analysis, Self-adjointness}. Academic Press, San Diego, 1980.

\bibitem[Re74]{Re74} G. Reents. \emph{Scattering of photons by an external current.} J. Math. Phys. \textbf{15}, (1974) 31--34.

\bibitem[RS20]{RS20} K.~Rejzner and M.~Schiavina. \emph{Asymptotic symmetries in the BV-BFV formalism}.  Commun. Math. Phys. \textbf{385}, (2021) 1083--1132. 

\bibitem[Ro70]{Ro70}  G. Roepstorff. \emph{Coherent photon states and spectral condition}. Commun. Math. Phys. \textbf{19}, (1970) 301--314.

\bibitem[Ru78]{Ru78} S.N.M. Ruijsenaars. \emph{On Bogoliubov Transformations II. The General Case}.
Annals of Physics \textbf{116}, (1978) 105--134.


\bibitem[Sh62]{Sh62} D. Shale. \emph{Linear symmetries of free boson fields}. Transactions of the American Mathematical Society \textbf{103},  (1962) 149--167.

\bibitem[SS65]{SS65} D. Shale and W. F. Stinespring. \emph{Spinor representations of infinite orthogonal groups}. Journal of Mathematics and Mechanics \textbf{14}, (1965) 315--322.

\bibitem[So23]{So23} J. Sorce. \emph{Notes on the type classification of von Neumann algebras}.  Rev. Math. Phys. \textbf{36},  (2024) 2430002.


\bibitem[So11]{So11}  R.V. Sol\'e. \emph{Phase Transitions}. Princeton University Press, 2011.

\bibitem[Sta81]{Sta81} A. Staruszkiewicz. \emph{Gauge invariant surface contribution to the number of photons integral}. Acta Phys. Pol. B \textbf{12}, (1981) 327--337.

\bibitem[St17]{St17} A.~Strominger. \emph{Lectures on the infrared structure of gravity and gauge theory}. Princeton University Press, 2018.


\bibitem[Vo03]{Vo03} M. Vojta. \emph{Quantum phase transitions}.
Rep. Prog. Phys. \textbf{66}, (2003) 2069--2110.

\bibitem[We65]{We65} S. Weinberg. \emph{Infrared Photons and Gravitons}. Phys. Rev. \textbf{140},  (1965)  B516.

\bibitem[Wi18]{Wi18} E. Witten. \emph{Notes On Some Entanglement Properties Of Quantum Field Theory}. arXiv:1803.04993.


\bibitem[Zi]{Zi} J. Zinn-Justin. \emph{Quantum Field Theory and Critical Phenomena}.
Oxford Science Publications, Fifth Edition, 2021.


\end{thebibliography}
\end{document}